\documentclass[aps,prd,amsmath,10pt,amssymb,twocolumn,showpacs,superscriptaddress,raggedbottom]{revtex4-1}

\usepackage{hyperref}
\usepackage{graphicx}
\usepackage{amsfonts,amsmath,amssymb}
\usepackage{color}

\begin{document}


\title{Tagging Spallation Backgrounds with Showers in Water Cherenkov Detectors}

\author{Shirley Weishi Li}
\affiliation{Center for Cosmology and AstroParticle Physics (CCAPP), Ohio State University, Columbus, OH 43210}
\affiliation{Department of Physics, Ohio State University, Columbus, OH 43210}

\author{John F. Beacom}
\affiliation{Center for Cosmology and AstroParticle Physics (CCAPP), Ohio State University, Columbus, OH 43210}
\affiliation{Department of Physics, Ohio State University, Columbus, OH 43210}
\affiliation{Department of Astronomy, Ohio State University, Columbus, OH 43210 \\
{\tt li.1287@osu.edu,beacom.7@osu.edu} \smallskip}

\date{November 23, 2015}

\begin{abstract}
Cosmic-ray muons and especially their secondaries break apart nuclei (``spallation'') and produce fast neutrons and beta-decay isotopes, which are backgrounds for low-energy experiments.  In Super-Kamiokande, these beta decays are the dominant background in 6--18 MeV, relevant for solar neutrinos and the diffuse supernova neutrino background.  In a previous paper, we showed that these spallation isotopes are produced primarily in showers, instead of in isolation.  This explains an empirical spatial correlation between a peak in the muon Cherenkov light profile and the spallation decay, which Super-Kamiokande used to develop a new spallation cut.  However, the muon light profiles that Super-Kamiokande measured are grossly inconsistent with shower physics.  We show how to resolve this discrepancy and how to reconstruct accurate profiles of muons and their showers from their Cherenkov light.  We propose a new spallation cut based on these improved profiles and quantify its effects.  Our results can significantly benefit low-energy studies in Super-Kamiokande, and will be especially important for detectors at shallower depths, like the proposed Hyper-Kamiokande.
\end{abstract}


\maketitle


\section{Introduction}

Astrophysical neutrinos can reveal the extreme physical conditions in their sources as well as new information about neutrino properties.  In the MeV energy range, the key targets are solar neutrinos and the diffuse supernova neutrino background (DSNB).  Solar neutrinos have been detected for half a century, yet there are still unanswered questions~\cite{GonzalezGarcia2008,Abe2011,Haxton2013,Aharmim2013,Antonelli2013,Renshaw2014}.  The upper limit on the DSNB flux is within a factor of a few of theoretical predictions~\cite{Beacom2010,Duan2010,Bays2012,Scholberg2012,Janka2012,Burrows2013,Zhang2015}.

Super-Kamiokande (Super-K) is a 50-kton water Cherenkov neutrino detector~\cite{Fukuda2003,Abe2014}.  Due to its large volume, low backgrounds, and long running time, Super-K has the best sensitivity to the high-energy, low-flux branches of the solar neutrinos and to the DSNB.

In Super-K, these measurements are background limited.  The dominant background in 6--18 MeV is the spallation background~\cite{Gando2003,Hosaka2006,Cravens2008,Abe2011}, which consists of beta decays from unstable isotopes produced by muons and especially their secondary particles~\cite{Li2014}.  These backgrounds are reduced by associating them with their muon parents, which can be difficult because some of these isotopes have long lifetimes (several seconds) compared to the muon rate ($\sim 2$ Hz)~\cite{Hosaka2006}.

In our first paper in this series~\cite{Li2014}, the only theoretical study of spallation in water, we calculated the average spallation yields in Super-K.  We compared the aggregate time profile and energy spectrum of spallation decays to Super-K measurements, finding agreement within uncertainties.  We showed that almost all isotopes are made by secondary particles, e.g., neutrons, pions, and gamma rays, instead of primary muons.

In our second paper~\cite{Li2015}, we showed for the first time that almost all spallation isotopes are made in muon-induced showers.  These showers have high densities of secondary particles; they extend only $\sim 5$ m along muon tracks, while the height of the Super-K detector is $\sim 40$ m.  Because showers can be detected through their Cherenkov light, this provides a new way to identify where a spallation isotope might be produced along the muon track.

Earlier, Super-K empirically found variations in the Cherenkov light intensity along muon tracks, and a correlation between the position of the peak and the spallation decay~\cite{Bays2012}.  They developed a new spallation cut based on the measured correlation.  Using this, they lowered the analysis energy threshold for the DSNB search.  However, the physical cause of the light variations and their correlation with spallation were unexplained.  Also, they did not apply this cut to their solar neutrino analysis.

Our finding that most spallation isotopes are made in showers explains Super-K observations, except one.  Their reconstructed muon light profiles are much broader and have much smaller amplitude than those expected from showers.  Here we show how this discrepancy can be explained by shortcomings of the Super-K reconstruction method, and how to improve it.  We explore applications of better-reconstructed profiles.  Our results should greatly benefit their solar neutrino and DSNB analyses.  Although we use Super-K as an example and attempt to model its main present features, our focus is more general.

For our calculations, we use the simulation package FLUKA (version 2011.2c.0)~\cite{Ferrari2005,Battistoni2007}.  It incorporates all the relevant physics for muon interactions in water.  Our physics choices for FLUKA are the same as in our previous papers~\cite{Li2014,Li2015}.  We simulate throughgoing muons vertically down the center of the Super-K detector; our results can be applied to more general cases.  The muon spectrum is shown in Fig.~1 of Ref.~\cite{Li2014}.  At the Super-K depth (2700 meter water equivalent), the average muon energy is 270 GeV~\cite{Tang2006,Li2014}.

One difference in our setup here is that our simulation region is the whole Super-K inner detector (ID), whereas in our previous papers we used only the fiducial volume.  The ID is a cylinder 33.8 m in diameter and 36.2 m in height~\cite{Fukuda2003}.  It is separated from the outer detector by opaque walls (including ceiling and floor), where photomultiplier tubes (PMTs) are mounted~\cite{Fukuda2003}.  The fiducial volume is an analysis region inside and smaller than the ID (22.5 kton versus 32 kton)~\cite{Fukuda2003}.  The PMTs collect light emitted in the whole ID, so we use it for our simulation volume.

This paper is organized as follows.  In Sec.~\ref{sec:light_profile}, we discuss the basics of shower physics and muon light profiles.  In Sec.~\ref{sec:reconstruction}, we review the Super-K reconstruction method and how to improve it.  In Sec.~\ref{sec:application}, we explore further applications of better-reconstructed shower profiles and quantify how much they could improve the spallation cut.  We conclude in Sec.~\ref{sec:conclusion}.


\section{Muon Cherenkov Light Profiles}
\label{sec:light_profile}

Relativistic charged particles in water emit Cherenkov light along their paths.  The Cherenkov photons propagate through the detector, occasionally getting scattered or absorbed.  Some of the photons reach PMTs and are detected.  The light intensity (number of photons per distance) emitted by a singly charged particle per unit distance is constant, independent of the particle type and energy~\cite{Jackson}.  The total number of photons is proportional to the energy deposited, and their arrival positions and times carry information about the event geometry.

When cosmic-ray muons pass through Super-K, they produce charged secondary particles, such as electrons and pions.  (With these generic terms, we typically mean $e^{\pm}$ and $\pi^{\pm}$; we separate $\pi^0$.)  The production and energy loss of secondary particles is prompt, much faster than muons crossing the Super-K ID ($\sim 100$ ns).  It is thus not straightforward to separate the Cherenkov light from cosmic-ray muons and their secondaries.

We call the Cherenkov light intensity along a muon track the muon light profile.  Its fluctuations reveal secondary production, because the light intensity from muons is constant.  To better describe the production of secondaries, we separate it into two steps: a primary muon directly produces daughter particles, and these daughter particles subsequently produce other secondary particles.

We quantify charged particles not by their number, but by the distance they travel, which is proportional to their Cherenkov light emission.  In FLUKA, charged particles are propagated by track segments.  The Cherenkov light emission from each segment is proportional to its length.

\subsection{Muon-produced Charged Particles}

The energy loss rate of a muon is
\begin{equation}
-\left\langle\frac{dE_\mu}{dx}\right\rangle = \alpha+\beta E_\mu ,
\end{equation}
where the brackets indicate averaging over distance~\cite{pdg}.  The $\alpha$ term is for ionization loss, and the $\beta E_\mu$ term is for radiative loss.  At the Super-K depth, where $\langle E_\mu\rangle=270$ GeV, $\alpha \simeq$ 2.9~MeV~cm$^{-1}$ and $\beta \langle E_\mu\rangle \simeq$ 0.7~MeV~cm$^{-1}$~\cite{Groom2001,pdg}.  

Ionization is muons losing energy by scattering bound electrons~\cite{pdg}.  We can further divide this term based on the energy of the outgoing electrons.  When their energy is small, this is restricted ionization; when it is large, this is delta-ray production~\cite{pdg}.  Restricted ionization loss is a continuous process, while delta-ray production is discrete interactions.  The boundary between these two cases is somewhat arbitrary~\cite{Battistoni2007}; we set it to be the electron Cherenkov threshold (kinetic energy 0.257 MeV).  The average total ionization energy loss for a muon that travels vertically through the ID is $\simeq 10$ GeV, with $\simeq 6$ GeV due to the restricted loss and $\simeq 4$ GeV due to delta-ray production~\cite{Li2014}.

Radiative processes include pair-production, bremsstrahlung, and photonuclear interactions~\cite{pdg}.  These are muons interacting with nuclei and producing electron-positron pairs, gamma rays, pions, and other mesons.  All of these processes have a large energy transfer for each discrete interaction (up to hundreds of GeV~\cite{pdg}), and the interaction rates are low.  The total radiative energy loss through the ID is $\simeq 3$ GeV.  All the energy that goes into radiative processes is carried by secondary particles, and mostly dissipates through ionization of these secondary particles.  Because of this, there is a near-constant relationship between energy loss and Cherenkov yield.  For the production spectra of daughter particles by muons in Super-K, see Fig.~7 of Ref.~\cite{Li2015}.

The fluctuation levels of the energy losses determine the features of muon light profiles.  The restricted ionization loss has negligible fluctuations along muon tracks, and among different muons.  The delta-ray production and radiative energy losses have large fluctuations.  Even though, on average, these terms are smaller than the restricted ionization, they can be much larger for individual muons.  We show how these features affect muon light profiles in the next subsection.

The energetic daughter particles from delta-ray production and radiative processes produce many secondary particles by inducing electromagnetic and hadronic showers~\cite{Li2015}.  A shower is a series of repetitive interactions where particles interact and multiply in number and decrease in energy.  When the average particle energy in the shower is too low to create new particles, particles range out by ionization.  An electromagnetic shower is mostly gamma rays producing electrons and positrons by pair production, and electrons and positrons producing gamma rays by bremsstrahlung.  A hadronic shower is mostly charged pions producing multiple charged and neutral pions, and neutral pions decaying to gamma rays and inducing electromagnetic showers.

\begin{figure}[t]
	\begin{center}
		\includegraphics[width=\columnwidth]{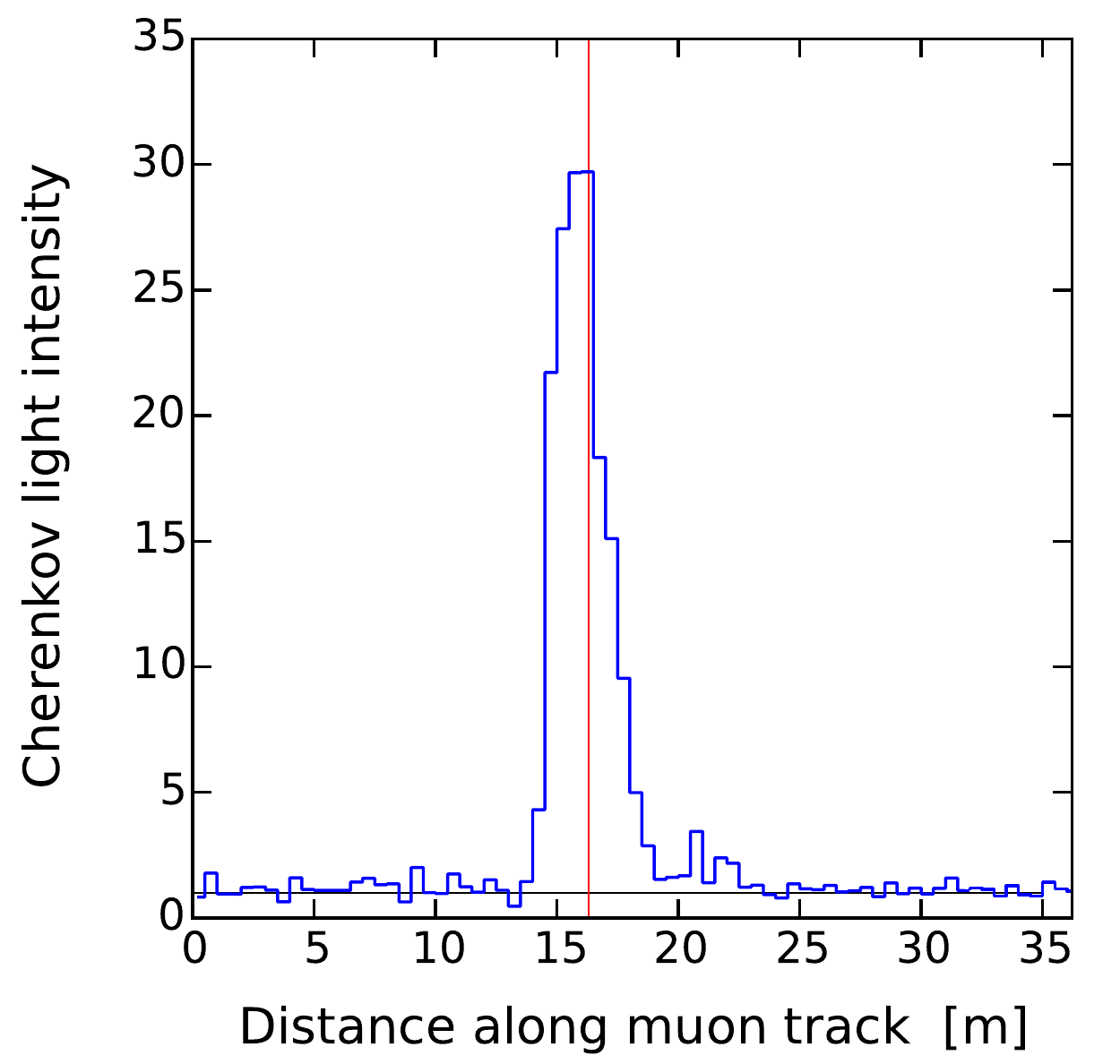}
		\caption{An example of a real (simulated) muon light profile.  The height is scaled to the light from a single muon.  The thin red line at 16 m indicates the peak position of the light profile.}
		\label{muon_profile}
	\end{center}
\end{figure}

\begin{figure}[t]
	\begin{center}
		\includegraphics[width=\columnwidth]{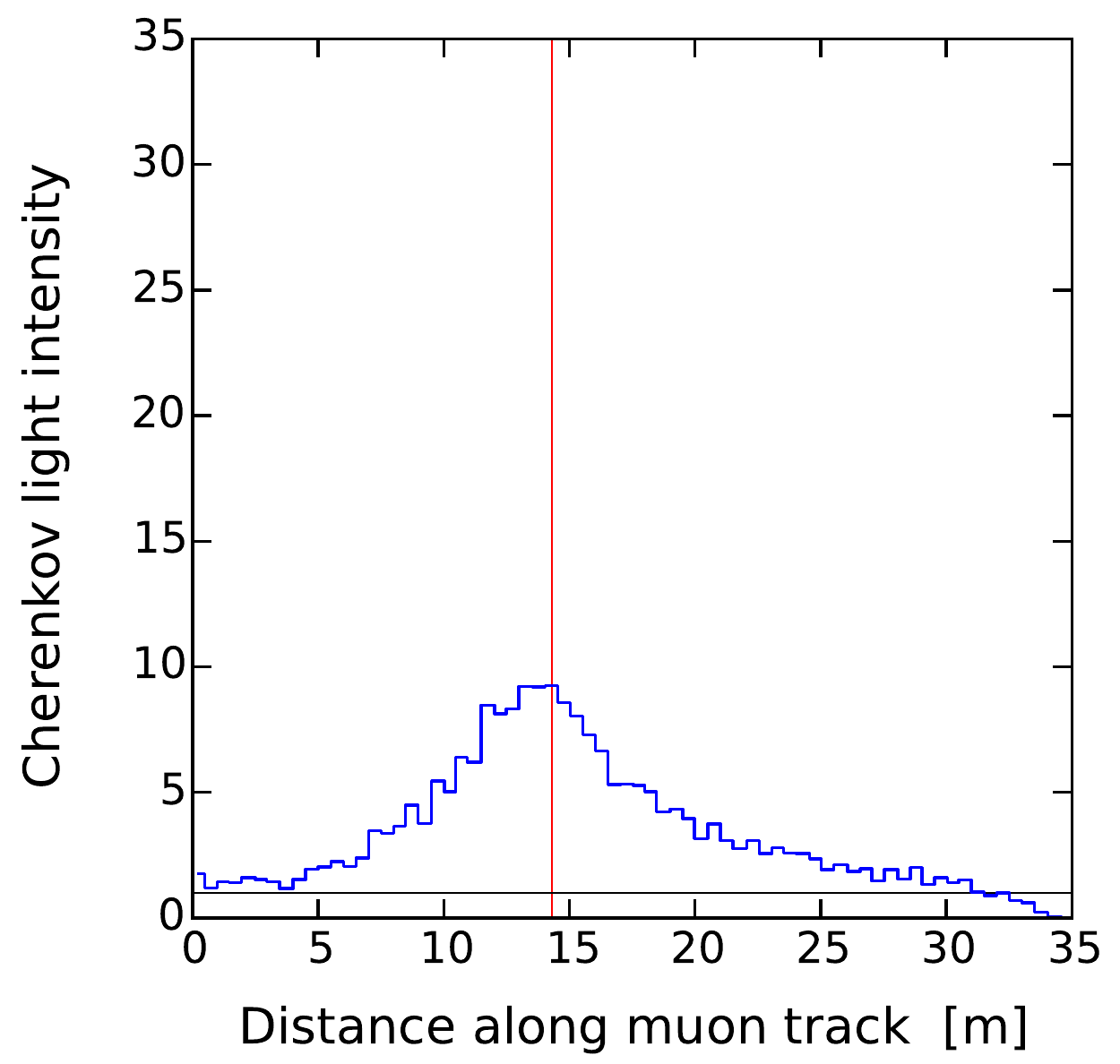}
		\caption{An example of a reconstructed muon light profile from Super-K (data from Fig.~2 of Ref.~\cite{Bays2012}).  The height is scaled to the light from a single muon.  The thin red line indicates the peak position of the light profile.  Compared to the profile in Fig.~\ref{muon_profile}, this peak is much wider and much shorter.}
		\label{sk_reconstruction}
	\end{center}
\end{figure}

An important energy scale for showers is the critical energy $E_c$; showers develop when the average particle energy is above it, and die out below it.  In water, $E_c \simeq 100$ MeV for electromagnetic showers~\cite{pdg} and $E_c \simeq 1$ GeV for hadronic showers~\cite{Li2015}.  Most showers in Super-K have energies $\simeq$ 1--300 GeV~\cite{Li2015}.  

Because most shower particles are energetic and have small deflections, showers look like long thin cylinders in real space (an example is shown in Fig.~1 of Ref.~\cite{Li2015}).  In terms of Cherenkov light production, the dominant contribution for either kind of shower comes from electrons near $E_c \sim 100$ MeV.  The average deflection of electrons is $\langle\cos\theta_z\rangle \simeq$ 0.8.  Showers in Super-K extend $\sim 5$ m in the longitudinal direction, and $\sim 10$ cm in the lateral direction (in hadronic showers, this can be $\sim 1$ m)~\cite{Li2015}.  For a detailed discussion of shower geometry in Super-K, see Sec.~IIC of Ref.~\cite{Li2015}.

Muon daughter particles with energy below $E_c$ do not induce showers.  The most common such particles are electrons.  Even though these electrons do not induce showers, they are important for our discussions because they emit Cherenkov light.  We refer to them as low-energy delta rays, i.e., electrons with energy above their Cherenkov threshold but below about 100 MeV.  We include in this a $\simeq 5\%$ contribution of low-energy electrons plus positrons from pair production.

Shower physics is the key to understanding how to use Cherenkov light to tag spallation backgrounds.  There are abundant electrons in electromagnetic and hadronic showers, so showers can be observed through their Cherenkov light.  There are few (many) pions and neutrons in electromagnetic (hadronic) showers, and they efficiently make isotopes.  By observing the light from a shower, Super-K could identify its position and thus the position where spallation decays might occur~\cite{Li2015}.  However, the light profiles observed by Super-K~\cite{Bays2012,Bays} look very different from what we expect from showers.  We take a closer look at this next.

\subsection{Real vs. Reconstructed Muon Light Profiles}

To utilize the correlation between showers and spallation decays, we need to know what showers look like in the detector.  In this subsection, we first study real muon light profiles, focusing on the shower shape.  By real we mean what reconstructed profiles would look like if every detected photon were reconstructed to its correct emission position; the profiles are simulated.  We then look at the differences between these profiles and those reconstructed from Super-K data.  

Figure~\ref{muon_profile} shows an example of a real muon light profile.  We simulate a vertical throughgoing muon in the ID and plot the total path length of relativistic charged particles relative to the muon path length.  This is equivalent to the Cherenkov light intensity in units of that from a single muon.  We adopt this unit because it is not affected by experimental effects, such as photon absorption and detector efficiency.  For this and similar figures, we use a bin size of 0.5 m, to be consistent with Super-K~\cite{Bays2012}.

The area under the curve is proportional to the muon energy loss.  A height of 1 corresponds to only restricted muon ionization energy loss.  Any height larger than 1 is due to ionization of additional charged particles, which are produced through delta-ray production or radiative processes.

Though just a single example, Fig.~\ref{muon_profile} is representative.  When muons are not showering, their profiles all look similar, with the relative intensity fluctuating between 1 and 2 due to the muon plus low-energy delta rays (the height is sometimes $< 1$ due to binning issues).  The average level is about 1.3, which corresponds to an extra energy loss of 0.5~MeV~cm$^{-1}$ due to low-energy delta rays.  The muon profile in Fig.~\ref{muon_profile} shows one energetic shower of energy $\simeq 15$ GeV.  (This example peaks near the center of the detector; showers can occur anywhere along the muon track.)  It is quite typical, extending $\sim 5$ m along the muon track and with a height of about 30.  For a shower of energy $E_0$, the peak height is typically 2--3 ($E_0$/GeV)~\cite{Li2015}.

The distance between the peak position of the light profile and a candidate signal event is used to determine the probability of the event being a spallation decay~\cite{Bays2012}.  In other words, Super-K keeps the peak position and discards information about the shape of the muon light profile.  In Sec.~\ref{sec:application}, we discuss how this can be improved.

The number of showers and the shower energies vary a lot for each muon~\cite{Li2015}.  The more energetic the daughter particle, the more rare it is.  Low-energy showers are thus more common than high-energy showers.  The average number of showers above 1 GeV per vertically throughgoing muon in Super-K is 0.4.  It is most common to have zero or one shower per muon.  Multiple showers along one muon track happen less than 10\% of the time.  

Super-K measured the variations of muon light intensity in Ref.~\cite{Bays2012}.  Though their approach is fairly general, it is not based on showers, in which electron deflections play a crucial role.  In the next section, we discuss how this affects muon light profile reconstruction.

Figure~\ref{sk_reconstruction} shows an example of a reconstructed muon profile from Super-K (another is Fig.~4.2 of Ref.~\cite{Bays}).  Their original $y$ axis is in units of number of photoelectrons detected.  We convert that to our units by using $7 \,\mathrm{p.e.} \simeq 1$ MeV~\cite{Ikeda}, which includes Cherenkov photon yield, photon absorption, and detector efficiency, etc.  This relation is not exact, but it does not affect our discussions.  

This light profile varies along the entire muon track, showing a prominent peak in the middle.  The full width of the peak is about 20 m and its height is about 10.  We estimate that the excess light corresponds to about 15 GeV, similar to the example shown in Fig.~\ref{muon_profile}.  Beyond 32 m, the falloff in intensity is probably because this muon track left the ID.

It is puzzling why the reconstructed profile looks like this.  Even though the shower in Fig.~\ref{muon_profile} is only one realization, it is representative of a shower of a similar energy.  The success of the current Super-K cut indicates that their average reconstructed peak position, at least for large muon energy losses, is quite robust.  In the next section, we explain the differences in the muon light profiles, and how to improve the Super-K reconstruction method to get accurate light profiles.


\section{Shower Reconstruction}
\label{sec:reconstruction}

It is important to understand the discrepancy between the real and reconstructed muon light profiles.  First, it is crucial to the proof that showers are the true cause of the variation in muon light intensity and that spallation isotopes are made in showers.  Second, as we pointed out in Ref.~\cite{Li2015}, the correlation function Super-K measured is not as sharp as the one we calculated using real shower profiles.  If Super-K could better reconstruct every shower, even if its energy is low or if it is accompanied by other showers, it would improve the efficiency of the spallation cut.

It is difficult to reconstruct the muon light profile.  During a shower, there are many charged particles emitting light at nearly the same time and position, but pointing in different directions, so it would be very difficult to resolve individual rings in the pattern of PMT hits on the wall.  Furthermore, the light from showers at specific locations must be separated from the continuous light from the muon itself and from low-energy delta rays.

In this section, we use our knowledge of showers, which are not mentioned in the Super-K paper, to examine the Super-K reconstruction method and its results.  We start by reviewing the setup of the equation that Super-K used for reconstruction.  Then we study in detail its properties and how its solutions are affected by physical and detector limitations.  We present ways to improve their method, demonstrating that we can reconstruct the muon light profile with high fidelity, including identifying showers and measuring their energy, position, and extent.

We first assume that the times of individual detected photons at a given PMT can be measured separately.  Then, in Sec.~\ref{sec:practical}, we discuss complications to that in Super-K and possible solutions.

\subsection{Super-K Reconstruction Method}

To reconstruct a muon light profile, Super-K performed backward fitting using individual PMT hits, without consideration of their correlations.  For each PMT hit (taken here to be one detected photon), they measured its position and time and solved for where along the muon track the photon was emitted.  They repeated this for all hits from one muon, and got the number of detected photons as a function of the emission position along the muon track, i.e., the muon light profile.

There is no confusion about which muon to associate a photon with, because photons from one muon arrive at the walls within $\sim 100$ ns, and the average time between muons is $\sim 0.5$ s (muon bundles are treated separately).  In addition, the short duration of the signal means that photons from unrelated low-energy decay backgrounds can be ignored.  Finally, because the water is so clear, the effects of light absorption and scattering are minimal.

\begin{figure}[t]
\begin{center}
\includegraphics[width=\columnwidth]{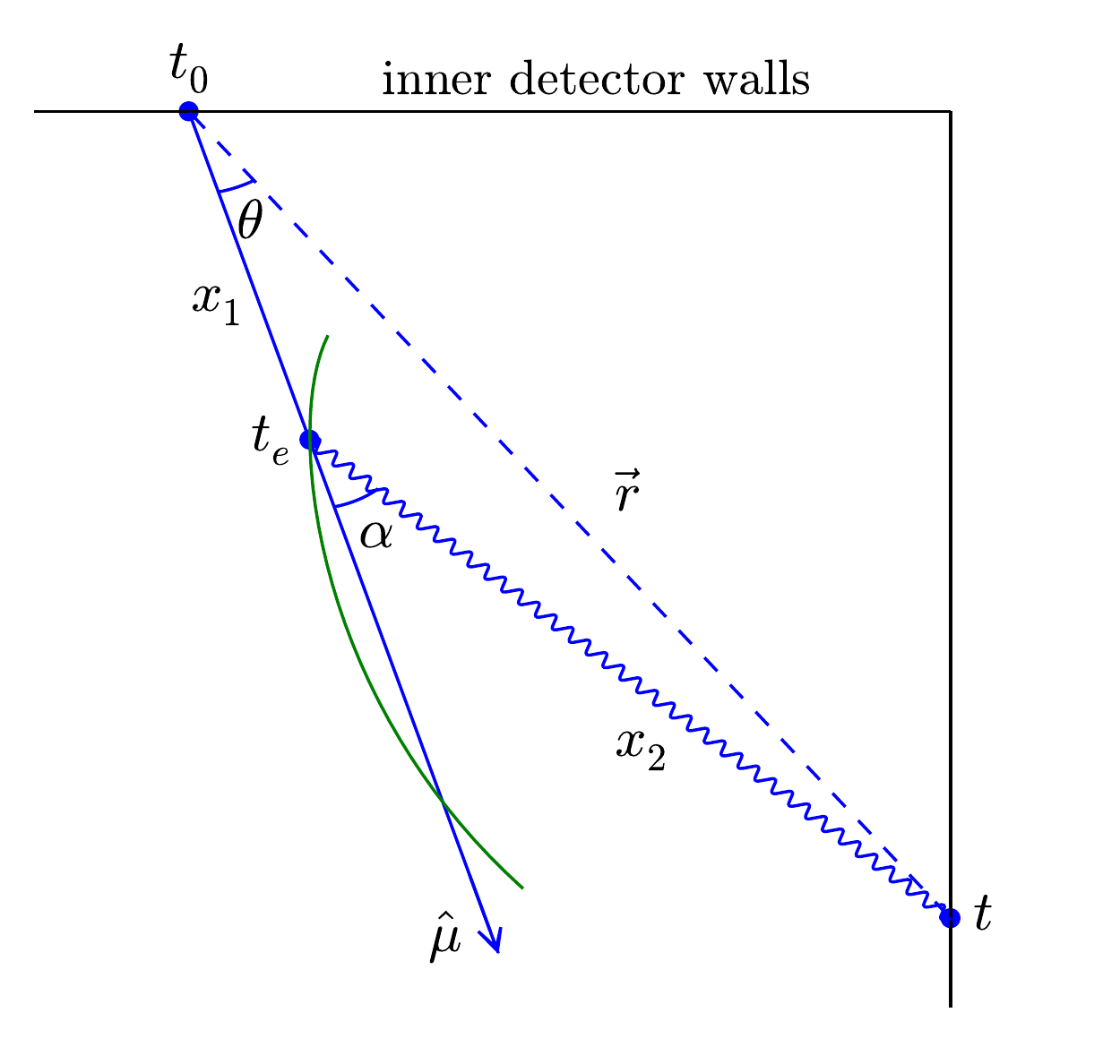}
\caption{Diagram for reconstruction of the muon Cherenkov light profile.  A photon is emitted from the muon track $\hat{\mu}$ at time $t_e$ and distance $x_1$, propagates a distance $x_2$, and hits a PMT at time $t$ and position $\vec{r}$.  The blue triangle (with corners marked by $t_0, t_e, t$) is described by Eq.~(\ref{eq:triangle}) and the green curve by Eq.~(\ref{eq:timing}); they cross at the solutions of Eq.~(\ref{eq:quadratic}), of which only one is marked.}
\label{sk_geometry}
\end{center}
\end{figure}

Figure~\ref{sk_geometry} illustrates the geometry.  A muon enters the ID at time $t_0$ and moves along the direction $\hat{\mu}$ with speed $c$.  A Cherenkov photon is emitted at time $t_e$, a distance $x_1 = c(t_e - t_0)$ along the muon track by either the primary muon or a secondary charged particle.  The photon propagates a distance $x_2$ to the PMT with group velocity $c/n_g$~\cite{Kuzmichev2002,Price2001,Badea2010} ($n_g=1.38$ for water).  The photon hits the PMT at time $t$, where its position is $\vec{r}$ relative to the muon entry point.  Because this method treats one detected photon at a time, in essence redrawing Fig.~\ref{sk_geometry} for each one, it is easy to accommodate muons at arbitrary positions and angles, as well as PMT hits on the ceiling or floor of the ID.  The azimuthal angle of each PMT hit in fixed detector coordinate is needed to define the plane of Fig.~\ref{sk_geometry}, after which it is not used.

In Fig.~\ref{sk_geometry}, the muon position at every instant, and thus $\hat\mu$ and $t_0$, is known because the entry and exit points and times are determined using inner and outer detector information.  The PMT hit time $t$ and its position relative to the beginning of the muon track, $\vec r$, are measured.  The angle $\theta$ is known immediately.  

The angle $\alpha$ can be obtained after solving for $x_1$ and $t_e$.  For photons emitted by primary muons, $\alpha$ is the Cherenkov angle $\alpha_0$, defined by the photon phase velocity via $\cos\alpha_0 = 1/n_{ph}$~\cite{Kuzmichev2002,Price2001,Badea2010}.  Its value in water is $\alpha_0=42^\circ$, with $n_{ph} = 1.33$.  Notice here that $n_g$ and $n_{ph}$ have similar values but are not equal.

The distance $x_1$ and time $t_e$ can be calculated as the joint solutions of two separate constraints, as illustrated in Fig.~\ref{sk_geometry}.  The spatial constraint is
\begin{equation}
x_2^2 = r^2 + x_1^2 - 2 x_1 r \!\cos{\theta} ,
\label{eq:triangle}
\end{equation}
which is satisfied by all points along the solid blue line of the muon track.  The time constraint is
\begin{equation}
x_1 + n_g x_2 = c (t - t_0) ,
\label{eq:timing}
\end{equation}
which is satisfied by all points along the green curve for which the total time --- accumulated as a muon for a distance $x_1$ and as a photon for a distance $x_2$ --- is $t - t_0$.  If $n_g = 1$, this curve would be an ellipse with foci at the points associated with $t_0$ and $t$; instead, it is a fourth-order polynomial, and we show a relevant section.

The joint solutions are defined by a quadratic equation in $x_1$, obtained by combining the constraints:
\begin{equation}
\Big[\frac{1}{n_g^2}-1\Big]x_1^2+2\Big[\vec{r}\cdot\hat{\mu}-\frac{c(t-t_0)}{n_g^2}\Big]x_1+\Big[\frac{c^2(t-t_0)^2}{n_g^2}-r^2\Big]=0 .
\label{eq:quadratic}
\end{equation}
This is the same as Eq.~(4.3) in Ref.~\cite{Bays}; we provide more details on the origin and solutions of this equation.  In principle, there could be zero, one, or two solutions.  Because every observed photon is emitted at some point on or close to the muon track, there should always be at least one real solution.  In Sec.~\ref{sec:solutions}, we show that numerical issues can make it appear that there are none.  The Super-K procedure is to measure the coefficients in this equation and solve for $x_1$, keeping all solutions that are real and fall into a reasonable range.

The Super-K reconstruction method is compatible with the properties of showers, because it is approximately true that all the light is emitted from a point moving along a straight line.  The longitudinal spread of particles at one instant of a shower is $\sim 0.1$ m.  For an electromagnetic shower, the transverse extent is also $\sim 0.1$ m; for a hadronic shower, it can be $\sim 1$ m.  Finally, typical muon track deflections are small, coincidentally $\sim 0.1$ m over the height of the ID.  As we discuss below, the typical resolution for shower reconstruction is of order a few meters, so these effects are negligible.

This method is powerful because it accounts for all of the observed Cherenkov light.  For relativistic muons, a photon is emitted at a fixed angle $\alpha=\alpha_0$ relative to the muon track; in this case, the emission point of the photon could also be obtained from a simpler linear equation, e.g., Eq.~(\ref{eq:projection_z}).  (An earlier Super-K paper~\cite{Desai2008}, on atmospheric neutrinos, reconstructed the muon light profile assuming that all emission was at the Cherenkov angle relative to the muon.)  However, for electrons, which can be significantly deflected, the angle $\alpha$ varies, and the full quadratic equation is needed.  The major shortcoming of this method, in either the quadratic or linear case, is that it neglects correlations between PMT hits, as discussed in detail in Sec.~\ref{sec:suggestions}.

As we have shown, the Super-K reconstructed light profiles are inconsistent with what we expect from a shower (or any known process).  Yet, at first glance, the Super-K reconstruction method looks correct.  To understand the differences between the real and reconstructed light profiles, we take a closer look at the nature of solutions of Eq.~(\ref{eq:quadratic}).

\subsection{Understanding the Reconstruction Solutions}
\label{sec:solutions}

To demonstrate how reconstruction works, we first simulate the Cherenkov light pattern that Super-K observes on the ID walls from muons and their secondaries.  We define the $z$ axis to point in the direction along the muon track.  For every PMT hit on the walls, Super-K can measure its position $z = \vec{r}\cdot\hat{\mu}$ and time $t$.  By light pattern, we mean the ($z$, $t$) plane filled by all the PMT hits from one muon and its secondaries.  The azimuthal angle for each PMT hit defines a plane for Fig.~\ref{sk_geometry}; when the light pattern is constructed, those azimuthal angles are discarded.

\begin{figure}[t]
\begin{center}
\includegraphics[width=\columnwidth]{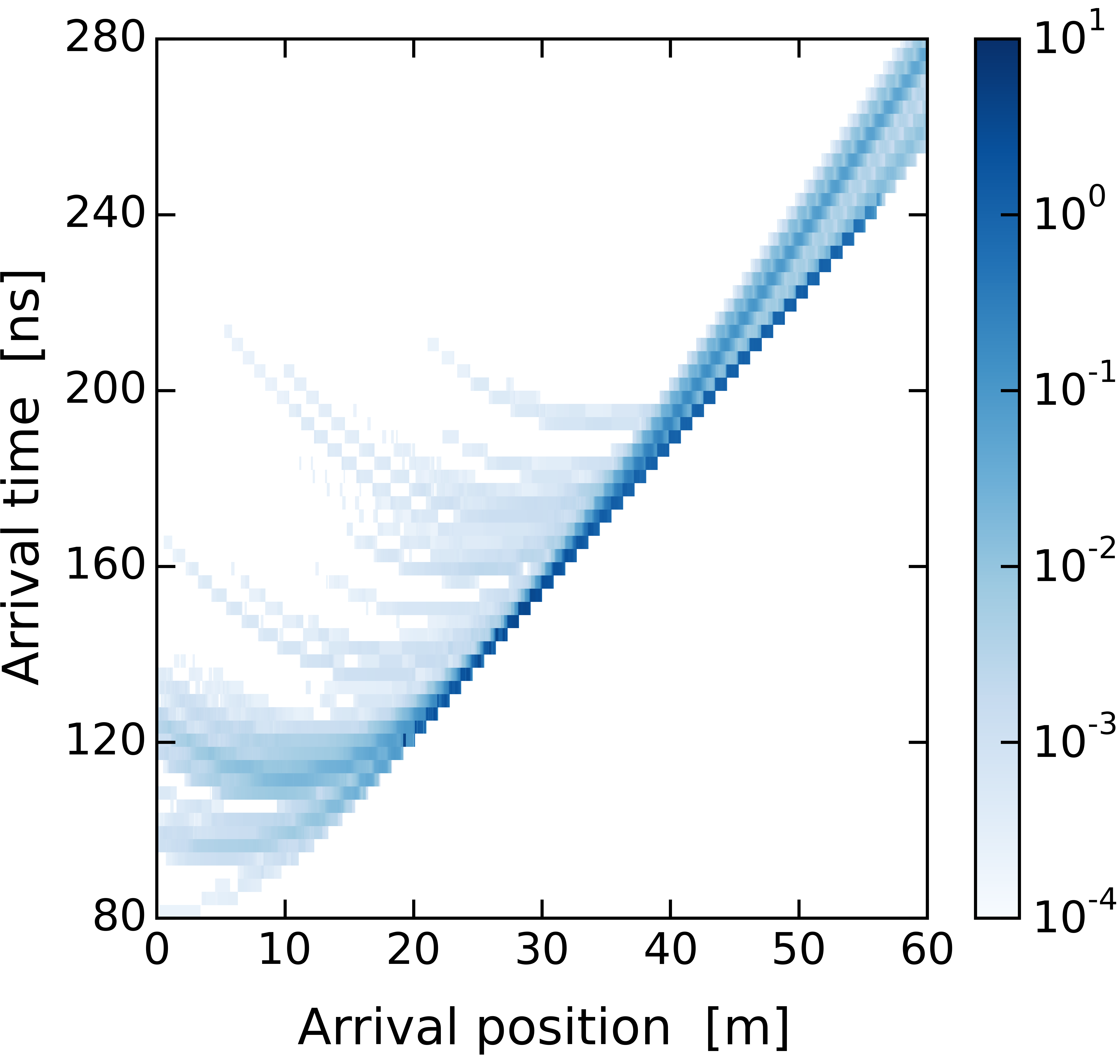}
\caption{The time and position ($z = \vec{r}\cdot\hat{\mu}$) pattern of Cherenkov light on the ID walls for one example of a muon and its secondaries.  We simulate an infinitely tall cylinder, such that all the photons hit the side walls instead of the ceiling or floor.  The arrival position range is thus larger than the height of the Super-K ID (36.2 m).  The intensity scale is approximate; the light from muons is about 1 on the scale.  The real muon light profile for this example is shown in Figs.~\ref{two_solution_comparison} and~\ref{reconstruction}.}
\label{wall_jet}
\end{center}
\end{figure}

To calculate the light pattern, we make some reasonable simplifications; a full study by Super-K will be needed to fine-tune the details.  We use a cylinder with the same radius (16.9 m) as the Super-K ID, but take it to be much taller, so that all light is collected on the side walls, instead of the ceiling or floor.  We take the muons to be vertically downgoing at the center of the cylinder, so that there is azimuthal symmetry on average.  For more general cases where the muon is tilted or shifted from the center, the appearance of Fig.~\ref{wall_jet} changes and we discuss it separately.  We use (m, ns) as the units, suppressing their display below.  The muon enters the ID at $(0, 0)$.

We generate and propagate Cherenkov photons geometrically with our own codes, ignoring light absorption and scattering.  For each track segment ($\sim 1$ cm) in FLUKA charged particle propagation, we effectively propagate light with intensity equal to its length along its Cherenkov cone from the midpoint of the segment.

The Super-K data are discrete in $z$ (all points on the surface of a PMT are taken to be the center of the PMT) but continuous in $t$ (though there is smearing due to time resolution).  For computational reasons, it is simpler for us to take $z$ to be near-continuous (bins of width 0.05 m) and $t$ to be discrete (bins of duration 3 ns).  For each $z$ value, we calculate $t$ and round it to the nearest bin.  If a segment is tilted away from the vertical, we uniformly distribute its light between the minimum and maximum $z$.  The discreteness of 3 ns in $t$, which is comparable to the timing resolution of the Super-K PMTs, is roughly equivalent to 1 m in $z$.  We discuss resolution further below. 

As a check, we generate the light patterns with the Cherenkov light propagation in FLUKA, including light absorption and scattering, and the results are consistent.  Furthermore, when we use the Super-K reconstruction method, we recover light profiles similar to theirs.

Figure~\ref{wall_jet} is an example of the light pattern produced by a muon and its secondaries.  The diagonal band with the highest intensity is due to the muon and the most forward electrons in one moderate shower, a couple of small showers, and some low-energy delta rays (which have $\langle \cos \theta_z \rangle \sim 0.85$).  The remainder of the intensity, less than 20\%, which arrives at given positions at later times, is due to significantly deflected electrons; that in the bottom left corner of the figure is due to very deflected electrons that produce light near the ceiling of the detector, where the muon cannot.

How much of this Cherenkov intensity (per bin of time and position) can be detected?  The first impression shows blocks of size 1 m, though a zoom-in reveals the 0.05-m bins; both have height 3 ns.  Because 1 m of a charged particle corresponds to $\sim 10^3$ PMT hits, Super-K should be able to detect some light from such blocks with intensities $\gtrsim 10^{-3}$ (or from 0.05-m bins with intensity $\gtrsim 10^{-2}$), assuming each is integrated over 3 ns.

The pattern in Fig.~\ref{wall_jet} from a muon (or any track segment along the $z$ axis) can be understood from simple arguments.  The setup is shown in Fig.~\ref{sk_geometry}, but here we assume a vertical central muon.  The photons emitted at a point $x_1$ have
\begin{equation}
z = x_1 + 16.9 \cot \alpha_0 = x_1 + 19,
\label{eq:projection_z}
\end{equation}
and
\begin{equation}
t = \left(x_1 + \frac{16.9\,n_g}{\sin{\alpha_0}}\right)/c = \frac{x_1}{c} + 118,
\label{eq:projection_t}
\end{equation}
where 16.9 (m) is the Super-K ID radius and $\alpha_0$ is the Cherenkov angle.  (For noncentral or nonvertical muons, these and the following expressions can be generalized by changing 16.9 to $r \sin\theta$.)  The offsets correspond to the position shift and time delay for light to reach the walls after the muon first enters the ID.  These equations combine to give the pattern
\begin{equation}
z = c\,t - 16.9 \left(\frac{n_g}{\sin\alpha_0}-\cot\alpha_0\right)
\label{eq:linear}
\end{equation}
in Fig.~\ref{wall_jet}.  This line is broadened to the left, into a band, due to finite detector time and position resolution.

The pattern in Fig.~\ref{wall_jet} from deflected electrons also has a characteristic shape.  For deflection by an angle $\theta_z$, the minimum and maximum $z$ for the propagated photons are $x_1 + 16.9 \cot (\alpha_0\pm \theta_z)$.  In a shower, there are many electrons in a short distance, some with very large deflections.  For light emitted at all angles from a single point $x_1$, the complete pattern is a hyperbola,
\begin{equation}
(z - x_1)^2 - \frac{c^2(t - t_e)^2}{n_g^2} = -16.9^2,
\label{eq:hyperbola}
\end{equation}
placed to the left of the line defined by the muon.  The lowest point of the hyperbola, $z_\mathrm{low}$, comes from light that travels perpendicular to the wall, which reveals the point of emission through $x_1 = z_\mathrm{low}$.  The right-hand side of the hyperbola is populated by light from electrons aligned close to the muon track; the left-hand side by light that is moving upward in the detector, due to very deflected electrons.  Because shower electrons are forward-peaked ($\langle \cos \theta_z \rangle \sim 0.8$), the intensity on the hyperbola falls as the electron deflection increases.  Outside the range of Fig.~\ref{wall_jet}, the Cherenkov intensity is nonzero but negligible.

For other muon positions or orientations, the appearance of Fig.~\ref{wall_jet} changes.  For a noncentral but vertical muon, the line from the muon light turns into a band.  In this case, photons emitted at the same point but at different azimuthal angles travel different distances to the wall.  Consequently, for a fixed $z$ value on the wall, it gets hit by photons emitted at different positions along the muon track, and they take different times to reach $z$.  For nonvertical muons, the width of the band varies because a nonvertical muon can be considered to be many small tracks of noncentral vertical muons.  These effects do not change our results because our reconstructions are based on Fig.~\ref{sk_geometry} and Eq.~(\ref{eq:quadratic}), which are fully general for each PMT hit.  We obtain good results for other muon positions and orientations.  In effect, for each range of azimuthal angle and height in detector coordinates, the pattern of PMT hits looks like that shown in Fig.~\ref{wall_jet}.
 
The key to understanding Fig.~\ref{wall_jet} --- and hence the reconstruction method --- is electron deflections.  We quantify these by the discriminant of Eq.~(\ref{eq:quadratic}),
\begin{equation}
\Delta = \frac{4\times16.9^2}{\sin^2\alpha} \left(\cos\alpha-\frac{1}{n_g}\right)^2,
\end{equation}
which determines the nature of the solutions (for central vertical muons).  If the phase and group velocities were identical, $\Delta$ would be zero for photons emitted by muons.  For these, $\Delta$ $\simeq 1$, much smaller than typical values for electrons.  To simplify the discussion, we approximate $\Delta \simeq 0$ for muon light.

Most electrons are quite forward ($\alpha \simeq \alpha_0$), so $\Delta$ is usually small; because measured PMT hits correspond to physical solutions, $\Delta$ must be positive.  When an electron is aligned with the muon, $\Delta = 0$ and the two solutions merge, corresponding in Fig.~\ref{sk_geometry} to the green curve from Eq.~(\ref{eq:timing}) being tangential to the blue muon line from Eq.~(\ref{eq:triangle}).  When an electron is deflected, $\Delta > 0$ and there are two real solutions, as in the example shown in Fig.~\ref{sk_geometry}.  As $\Delta$ increases, the two real solutions split further apart.  Although both are physical, one is correct and one is not, and these cannot be distinguished on an event-by-event basis.  The Super-K reconstruction method keeps both solutions if they are within a reasonable range, which leads to a problem of {\it overcounting} PMT hits (about 35\% in the example of Fig.~\ref{wall_jet}).

The typically small discriminant amplifies the effects of the detector time and position resolution.  The measured $\Delta$ can become negative, corresponding to no real solutions, due to shifts in the measured quantities used in the coefficients of Eq.~(\ref{eq:quadratic}).  In Fig.~\ref{sk_geometry}, the green curve would be slightly displaced from the blue muon line, with no crossings; in Fig.~\ref{wall_jet}, the PMT hits would be slightly to the right of the line defined by the muon.  The Super-K reconstruction method discards such cases, which leads to a problem of {\it undercounting} PMT hits (about 40\% in the example of Fig.~\ref{wall_jet}).

These numerical problems also mean that true solutions depend sensitively on the measured values of $(z, t)$.  In Fig.~\ref{sk_geometry}, the near-straightness of the green curve means that slight movements of it or the blue muon line lead to large changes in the solutions.  In Fig.~\ref{wall_jet}, it is difficult to separate hyperbolas with different emission points $x_1$ by looking at their right-hand sides, where the Cherenkov intensity is greatest.  More quantitatively,
\begin{equation}
x_1(z,t) \simeq x_1(z_0,t_0)+ \frac{2.6}{\sqrt{\Delta}} (ct - z) (\delta z - c\delta t),
\label{eq:derivative}
\end{equation}
where $\delta z$ and $\delta t$ describe how incorrect the values of $(z, t)$ are.  When $\Delta$ is small, the error term scales as $\sim 20 / \sqrt{\Delta} \sim 0.5|\cos\alpha - \cos\alpha_0|^{-1}$ m, which can be several meters for typical shower electrons.  This is significantly larger than the PMT position or time resolution because of the near-cancellation in $\Delta$.  (As we show in Sec.~\ref{sec:improvC}, when $\Delta$ is near zero, we can solve a linear equation instead of the quadratic.)  Importantly, this tells us that {\it the best reconstructions come from the worst electrons}, i.e., those with the largest deflections and $\Delta$ values.

The nature and precision of the solutions can also be affected by the presence of hadronic showers.  For these, transverse displacements of some shower particles can be $\sim 1$ m from the muon track, especially for large showers.  This means that $\Delta$ values are shifted compared to the case with no deflection.  This can be seen from the fact that there are more negative $\Delta$ values and they can have larger absolute value, compared to photon hits from electromagnetic showers.  The effect is small enough that we can ignore it here, but large enough that it could help identify hadronic showers, which are rare but which produce nearly all isotopes.

With a better understanding of the solutions, we now have insights as to why the Super-K reconstructed profile looks very different from real shower profiles.  We next consider how to improve their method.

\subsection{Improving the Reconstruction Method}
\label{sec:improvC}

The exploration in Sec.~\ref{sec:solutions} reveals how the Super-K reconstruction method works, as well as three improvable limitations.  First, when $\Delta$ is large, taking both solutions includes wrong information and overcounts PMT hits.  Second, when $\Delta$ is negative due to detector resolution, taking zero solutions ignores correct information and undercounts PMT hits.  Third, when $\Delta$ is small, the sensitivity of the solutions to detector resolution dilutes better information.  

These limitations result in reconstructed muon light profiles with distorted shapes and inaccurate shower energies.  When the shower energy is small, the current method might not be able to localize the shower.  Multiple showers cannot be resolved either.  

Our goal for improving the reconstruction method is to get an accurate muon light profile.  This includes locating the correct shower peak position, getting the correct shower shape and shower energy, and resolving multiple showers.  We improve the Super-K method by addressing its three limitations.  First, when $\Delta$ is large, we show how to select the better solution (Improvement 1).  Second, when $\Delta$ is negative, we show how to repair it and recover a solution (Improvement 2).  Third, when $\Delta$ is small, we show that, though these solutions help reconstruct the complete light profile, it is best to set them aside when defining the showers (Improvement 3).

\begin{figure}[t]
\begin{center}                  
\includegraphics[width=\columnwidth]{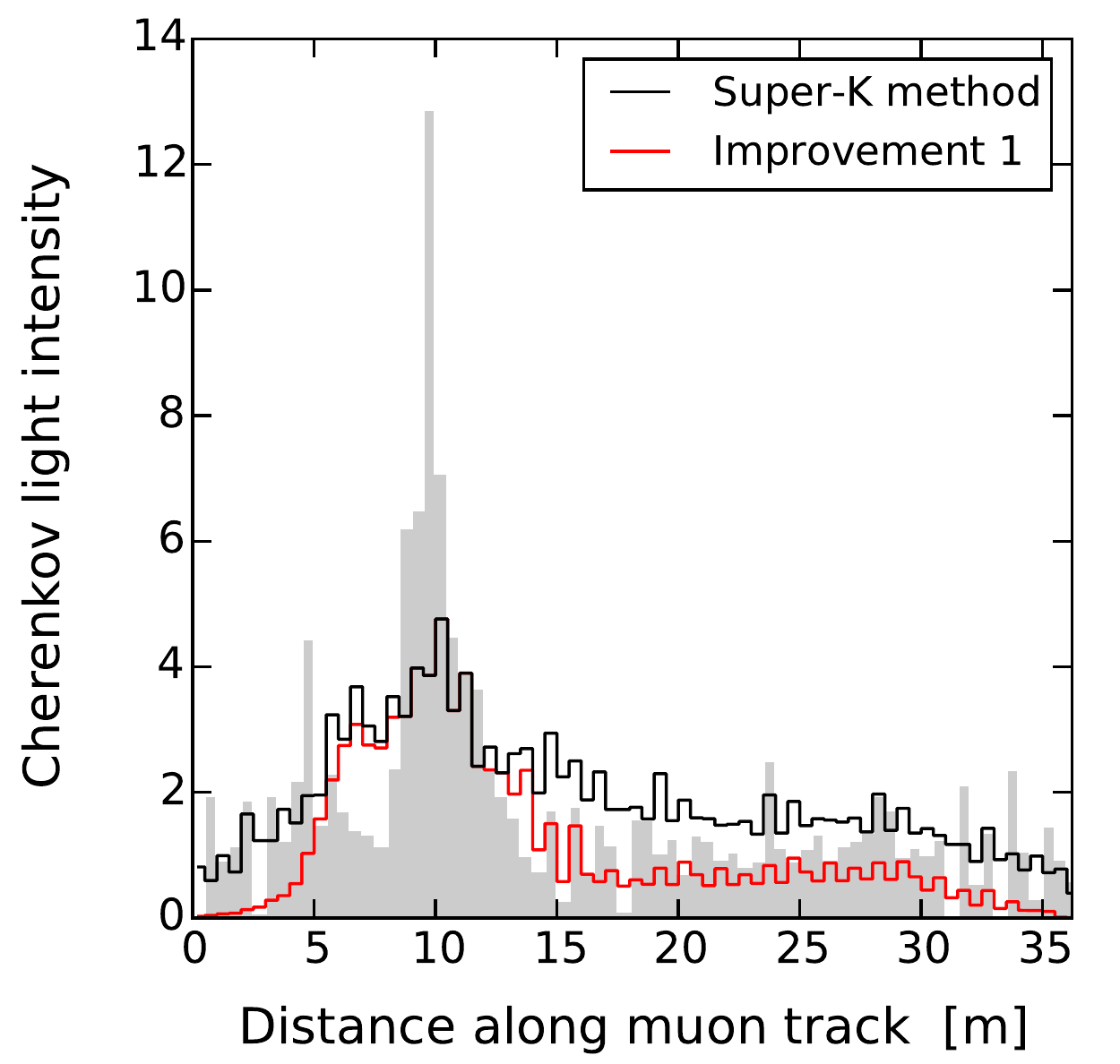}
\caption{Shower reconstruction with refinement of real solutions.  The gray shaded shape is the real (simulated) light profile used to produce the example light pattern in Fig.~\ref{wall_jet}.  The black line is the result of the reconstruction using the Super-K method.  For the red line, we keep only the better one of the two solutions.}
\label{two_solution_comparison}
\end{center}
\end{figure}

Figure~\ref{two_solution_comparison} shows (in gray shade) the Cherenkov light profile from a simulated muon and its secondaries.  Using this specific example, we calculated the $(z, t)$ data shown in Fig.~\ref{wall_jet}; here we use that data as if they were observed, attempting to reconstruct an accurate light profile from it.  

In this example, the total muon energy loss is 11 GeV.  There is a medium-sized shower (about 4 GeV, as can be seen from its area) located near 10 m, a smaller shower near 5 m, and possibly some smaller ones further along the muon track.  These are all quite typical in appearance, with the smaller showers being harder to recognize.  These particular showers are electromagnetic; there are harmless isotopes produced by gamma rays at around 5 m and 30 m.  As explained before, the light outside the shower regions is from the muon and low-energy delta rays.  It may look like there are larger fluctuations in the muon and delta-ray light than in Fig.~\ref{muon_profile}, but this is only because the overall $y$ scale is smaller due to this shower being smaller.  

We choose this example because the biggest shower has only moderate energy and because there are two showers close to each other.  It is a good test of how well the reconstruction works, both in terms of getting the correct shape of the largest shower and of resolving the small showers.

Figure~\ref{two_solution_comparison} also shows (black line) the result obtained when we use the Super-K reconstruction method.  The largest shower is found at the right position, which is why the new Super-K spallation cut works, as shown in Fig.~12 in Ref.~\cite{Li2015}.  However, the shape of this shower is badly smeared.  The area under the black line is comparable to that in the gray shade, but this is an accident, because the Super-K method overcounts and undercounts by roughly equal amounts, as noted.  The small showers are not resolved.

Figure~\ref{two_solution_comparison} also shows (red line) the result if we improve the Super-K method by keeping only one solution when there are two (Improvement 1).  We first run the Super-K reconstruction method, and record the peak position from the reconstruction.  We then run the reconstruction a second time, keeping only the solutions closer to the peak.  The red line agrees with the black line near the peak, as expected, but is lower elsewhere because it is not including wrong solutions and overcounting.  Though it defines the showers better, it is still significantly broader than it should be.  We note that when the Super-K reconstruction method produces a wrong peak, due to multiple showers or for small showers, Improvement 1 can reinforce it, but does not cause the problem.  Improvement 3 can fix the problem because ``fake'' showers would not have many deflected electrons, as shown in the Appendix.

Next we consider the data for which $\Delta$ is slightly negative due to detector resolution (Improvement 2).  Typical values are at worst $-30$ m$^2$ (nominal resolution $\simeq 3$ m [Eq.~(\ref{eq:derivative})]), corresponding to the level expected from detector resolution.  We reconstruct these cases by setting $\Delta = 0$, as would be appropriate for light from the muon or very forward electrons.  Because the angle of the particle is known, the quadratic equation reduces to a linear equation, e.g., Eq.~(\ref{eq:projection_z}).  We use the linear equation only when $\Delta = 0$ (or is reset to be), as the quadratic equation is less sensitive to numerical problems from detector resolution except for the smallest positive values of $\Delta$.

\begin{figure}[t]
\begin{center}                  
\includegraphics[width=\columnwidth]{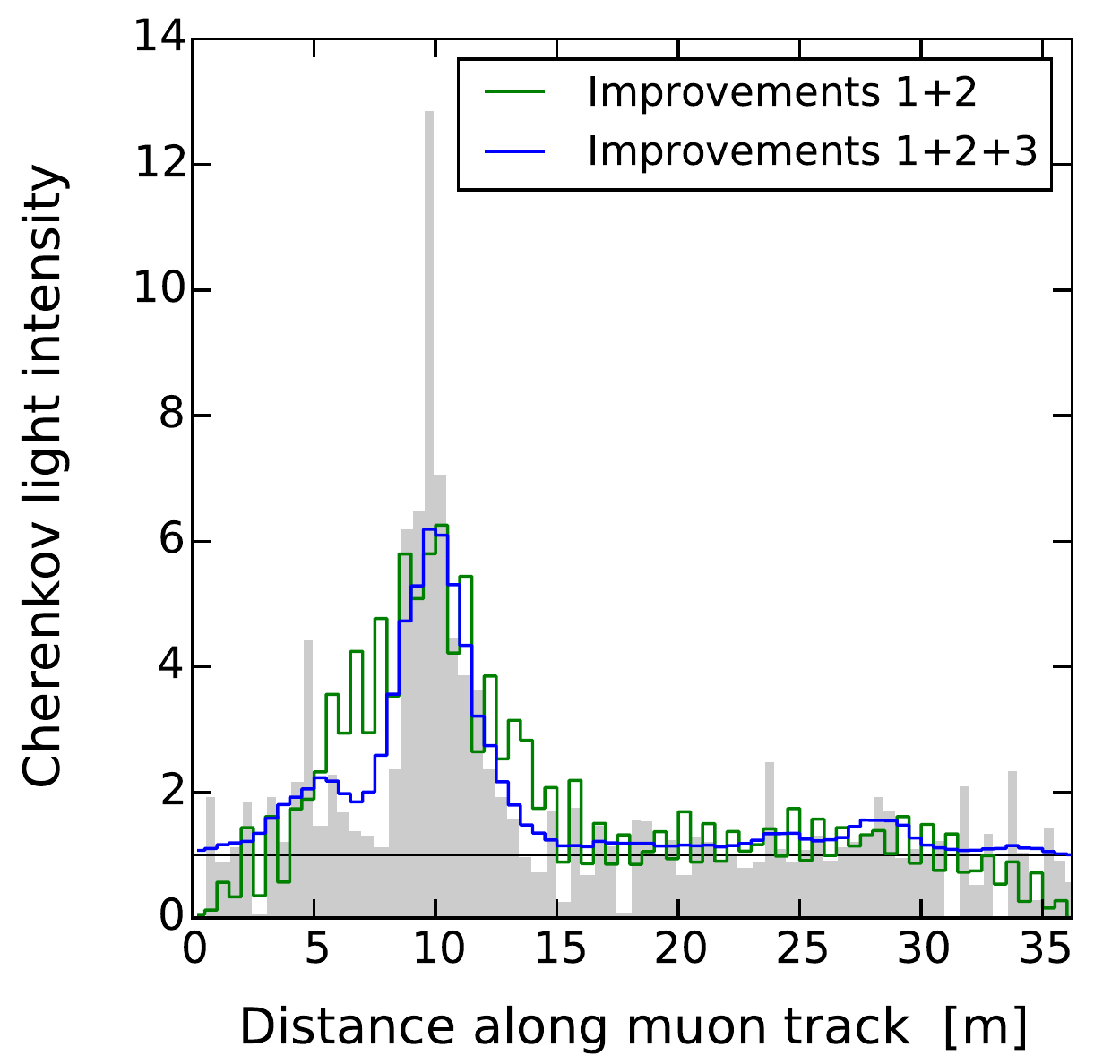}
\caption{Shower reconstruction with two more improvements.  The gray shaded shape is the real (simulated) light profile.  For the green line, we repair unphysical data to recover muonlike solutions that were previously discarded.  For the blue line, we select only the light from electrons with large deflections to focus on reconstructing only the showers; we add 1 to account for the muon light.}
\label{reconstruction}
\end{center}
\end{figure}

Figure~\ref{reconstruction} shows (green line) the result when we also recover these formerly discarded solutions (all of our improvements of the reconstructed muon profile are cumulative.)  This profile is a better match in the peak and especially the baseline to the input in the gray shade than even the red line in Fig.~\ref{two_solution_comparison}.  There is no longer undercounting of light from undeflected particles, such as the muon itself.  Importantly, the area under the green line now matches that in the gray shade.  However, the shower peak is still too wide, and no secondary showers are identifiable.  Because the green line traces the nonshowering part of the light profile well, it can be used to estimate the energy of the largest shower from its area above the baseline; at this stage of refinement, imperfect precision increases the width of the shower and decreases its height, but conserves its area.

Finally, we focus on the photons that provide the most precise information on the positions of showers (Improvement 3).  From Eq.~(\ref{eq:derivative}), these are the ones with large $\Delta$.  We choose $\Delta > 100$ m$^2$, corresponding to a nominal resolution of 2 m [Eq.~(\ref{eq:derivative})].  This comes at a price of keeping only $\sim 10\%$ of the PMT hits, corresponding to $\sim 1$ GeV.  We solve the quadratic equation, keeping only the solution closest to the shower peak as determined with the Super-K method.  We correct the normalization of the blue line by adding a constant baseline of 1 for a muon and by setting the shower energy above the baseline to match that of the green line.

Figure~\ref{reconstruction} also shows (blue line) the result obtained using only the most deflected electrons.  The agreement with the input shown in the gray shade is excellent.  Compared to previous results, it is much narrower, localizing showers better.  Only this method clearly defines multiple showers.  We added in the muon baseline to facilitate comparison, but the underlying method ignores the light from the muon and most of its low-energy delta rays.  That is, it focuses on the light in showers, where nearly all the isotopes are made.

We show all three improvements in this order to best explain the physics.  In practice, the first step is Improvement 3, which is to pick out the most deflected electrons.  Next is Improvement 1, which is to select only one solution for each PMT hit (for the most deflected electrons).  Lastly, one can follow Improvement 2 to get the correct total energy, then scale the profile from Improvement 3+1 to the correct energy.  However, one can also skip Improvement 2, and get the total energy simply by counting the total number of PMT hits, then scaling the profile to the correct energy.

We have demonstrated that the Super-K reconstruction technique can be significantly improved, leading to an accurate muon light profile, even when there are multiple showers.  However, there are some complications regarding practical implementation, which we discuss next.



\subsection{Towards Practical Implementation}
\label{sec:practical}

So far we have assumed that Super-K can measure the position and time of each detected photon within the precision noted in Sec.~\ref{sec:solutions}.  However, this is not always possible with the present electronics.  Here we discuss the implications and how to achieve the aim of shower reconstruction anyway.

A PMT hit is the basic observable and must be defined carefully.  In Super-K, it is defined by the total number of detected photons within a time window and the time of just the first photon.  The number of detected photons is determined by the accumulated charge of the photoelectrons produced.  In Super-K, the time window is $\sim 400$ ns (Michael Wilking, private communication; Michael Smy, private communication).  As assumed above, the position and number of detected photons are well defined, but the individual times are not, which reduces the available information.

This effect is important for reconstructing muon-induced showers in Super-K.  The light yield of a vertical throughgoing muon is high, corresponding to several detected photons per PMT, and more if there are large showers.  As shown in Fig.~\ref{wall_jet}, the light from the muon always arrives at a given PMT before that from a shower.  Because of this, most PMTs lose the timing information on the light from showers.  Despite this, Super-K found reasonable reconstructions in Ref.~\cite{Bays2012}, where they weighted the solutions corresponding to each PMT by the total number of detected photons.

Much of the data needed for reconstruction are not affected by this limitation.  The key is to identify cases where light from the muon does not reach the PMTs.  The most significant reason is due to geometry.  For vertically downgoing muons, the most common case, their Cherenkov light cannot reach the PMTs in roughly the top half of the detector.  This can be seen in Fig.~\ref{wall_jet}; the height of the ID is 36.2 m, and the muon light begins only at a depth of 19 m.  We emphasize that the PMTs in the top half of Super-K can detect photons from showers anywhere in the detector.  Indeed, the further the direction of the shower light is away from that of the muon light, the better.  Another reason is due to fluctuations.  Some PMTs that could have been reached by the muon Cherenkov light will not be triggered, and these will properly register late-arriving light.

To check the effects of the timing limitations, we constructed a second simulation, which is a more faithful representation of Super-K.  (We do not use this simulation for our main results because it complicates the discussion of the underlying physics.)  The simulated region follows the true Super-K ID geometry with the ceiling and the floor.  Individual PMTs are mounted on the ID walls with realistic sizes and spacing.  We record each photon hit with its total charge and first-hit time, as opposed to treating photoelectrons individually.  For the reconstruction, we repeat the Super-K method and our improvements.  For Improvement 3, not only do we select the photons with large deflection, but also hits on the ceiling and in the top half of the detector, where no muon light is expected.  Our reconstructed profiles reasonably trace the true profiles and can pick out showers occurring even near the bottom of the detector.  Thus we are confident that the properties of the PMT electronics will not significantly affect our results.

Longer term, the ability to reconstruct showers could be improved by installing new electronics that allow for pulse-shape discrimination, or at least enough information to separate detected photons that arrive a few tens of ns apart.  New technologies for Cherenkov light detection with excellent position and time resolution are extremely promising for improving shower reconstruction~\cite{Anghel2014,Anghel2015,Oberla}.

In the near term, the most promising possibility is to go beyond the framework of the Super-K reconstruction method and our improvements, and take advantage of the ideas proposed in the following subsection.


\subsection{Towards Better Reconstruction Methods}
\label{sec:suggestions}

The Super-K reconstruction method, including our improvements, works reasonably well.  However, it has fundamental shortcomings.  It neglects the correlations between different photons from the same charged particle, i.e., the Cherenkov ring pattern.  It neglects the correlations between different electrons emitted from the same position, i.e., the shower angular distribution.  And it neglects the correlations between electrons emitted from different positions, i.e., the shower longitudinal profile.

Better methods should be possible.  Here we sketch three promising ideas, each for a different energy range; it may be possible to combine them.  A good reconstruction needs only to provide the number of relativistic charged particles accompanying the muon, and some information about their angular distribution, each in bins of size $\simeq 0.5$ m along the muon track.  The muon and the shower each produce a lot of light, $\sim 7000$ PMT hits per GeV, which provides a lot of information for such modest goals.  We have had encouraging conversations with Super-K collaborators about specific codes that could be adapted to this purpose, such as fiTQun~\cite{Patterson2009,Wilking2013} (private communication, Michael Wilking) and MS-fit~\cite{Pik} (private communication, Michael Smy), if a pure enough sample of hits can be obtained.

To exploit the correlations between photons in the same Cherenkov ring, one must connect the solutions from separate PMTs.  Consider a vertically downgoing muon passing through the center of the detector (the considerations generalize).  The Cherenkov ring from the muon is a circle of uniform intensity moving down the detector walls.  Charged particles in a shower are arrayed in a small, thin, concave bunch centered on the muon.  The light from forward electrons adds to that from the muon, but the light from each deflected electron makes a tilted ellipse that intersects the circle from the muon at only two points.  When we fit only one PMT hit at a time, it is as if we are azimuthally averaging, turning these ellipses into broad circular bands that blend with the light from each other and that from the forward particles.

For showers of small energy, which are the most important in terms of the frequency of isotope production, it should be possible to simultaneously fit the Cherenkov rings of all charged particles, or at least the most deflected ones.  This method may also work for low-energy delta rays.

To exploit the correlations between electrons at the same position, one must take into account the angular distribution of shower particles.  Their light follows hyperbolas described by Eq.~(\ref{eq:hyperbola}) and clearly visible in Fig.~\ref{wall_jet}.  We emphasize that the Super-K reconstruction method, even with our improvements, does not exploit these hyperbolic patterns, which is clearly a missed opportunity.

For showers of intermediate energy, it should be possible to fit the portions of the hyperbolas in Fig.~\ref{wall_jet} that can be separated from the light from the muon track.  The angular distribution of electrons at a given point along the muon track determines how the intensity varies along the hyperbola.  It is probably adequate to focus on the integral of this intensity, which reveals the number of sufficiently deflected electrons at each position.

To exploit the correlations between electrons at different positions, one must take into account the longitudinal shower profile.  At present, the values of the reconstructed light from each distance bin along the muon track are independent.  This allows fluctuations between different bins that are larger than the intrinsic ones, which likely increases the apparent width of showers. There should be a way to enforce consistency between the values reconstructed for nearby bins.

For showers of large energy, it should be possible to do forward fitting with a template for a shower of unknown energy and position along the muon track, assuming something about the average angular distribution of shower particles.  As shown in Fig.~6 in Ref.~\cite{Li2015}, the intrinsic shower fluctuations at 100 GeV are minimal and those at 10 GeV are moderate; this method may work to even lower energies.

With these or other new methods based on the physics of showers, we are confident that the quality of the reconstructed light profiles can be significantly improved, resolving much smaller showers and multiple showers per muon.  This will allow much sharper cuts on spallation isotopes.


\section{Muon Profile Likelihood}
\label{sec:application}

Our goal is to improve spallation background rejection in Super-K, i.e., the separation of spallation decays from neutrino signal events.  Currently, Super-K uses spallation likelihood functions that take variables describing a candidate signal event and return a probability that it is a spallation decay.

So far our discussions have been within the framework of the Super-K likelihood function for their DSNB analysis~\cite{Bays2012}, which is based on finding the peak position of a muon light profile.  Our work in previous sections shows how to measure this peak better.

In this section, we propose a new framework.  We build a spallation likelihood function based on our faithfully reconstructed shower profiles.  We quantify its improvement to Super-K spallation cuts.

There is an important distinction between constructing a likelihood function and applying it.  When constructing a likelihood, one always knows which primary muon made a particular spallation isotope, whereas when applying a likelihood, one does not know which muon to associate a particular event with.  We explain the first part in Sec.~\ref{sec:A} and the second in Sec.~\ref{sec:B}.

\subsection{Spallation Likelihood Functions}
\label{sec:A}

In Super-K, solar neutrinos have a low event rate.  Intrinsic radioactivity backgrounds dominate at low energy ($<$~6~MeV); spallation backgrounds dominate at high energy (6--18~MeV).  Both neutrino events and radioactive backgrounds are uncorrelated with cosmic-ray muons, and we refer to them as random events.

A Super-K spallation likelihood $\mathcal{L}(C,M)$ evaluates how likely it is that a candidate event ($C$) is correlated with a muon ($M$).  The larger $\mathcal{L}$, the more likely that this $C$ is made by this $M$, i.e., is a spallation decay.  Otherwise, it is likely an uncorrelated random event.  (Below, we directly define likelihood functions; Super-K analyses use the logarithm of the likelihood, which is equivalent.)

A good likelihood function reflects the physics of how muons make spallation isotopes.  (Though well motivated on general grounds, the Super-K likelihood functions are empirical.)  There are several steps to build this function.  First, one picks variables describing a candidate event that are statistically different for spallation decays and random events.  Some obvious choices are the differences in time and transverse position between a candidate event and its parent muon as well as the muon energy loss.  A basic assumption that Super-K adopts, which we keep, is that these variables are independent, i.e., that the likelihood function can be factorized.

Second, one selects a spallation decay sample along with their parent muons.  In simulation, this is easy.  In practice, Super-K selects candidate events that are close to muons in time and space, and that have high energy (to avoid radioactive backgrounds).  This is sufficient to select an almost pure spallation sample, due to their high rate.

Third, one selects a random sample with uncorrelated muons.  In simulation, we simply produce candidate events that are uniform in time and space, and randomly pair them with muons.  In practice, Super-K pairs candidate events with muons that follow, instead of precede, candidate events in time.

Finally, one builds every component in a likelihood function.  For each variable, one measures the distributions of this variable for the spallation event sample and the random event sample.  Then, each likelihood component is the ratio of the distributions of the spallation sample relative to the random sample.

The first likelihood function that Super-K developed, which is still used for solar neutrino analyses~\cite{Hosaka2006}, is
\begin{equation}
\label{eq:solar}
\mathcal{L}_\mathrm{flat} = F_t(t)\cdot F_l(L_\mathrm{trans})\cdot F_q(Q_\mathrm{res}) .
\end{equation}
Here $t$ and $L_\mathrm{trans}$ are the time difference and the transverse distance between a spallation decay and a muon track, and $Q_\mathrm{res}$ is the radiative energy loss of the muon (measured by subtracting the light of a minimum-ionization muon from the total).   We can understand the behaviors of these components based on shower physics.  $F_t$ decreases due to the exponential decays of spallation isotopes.  $F_l$ decreases due to the exponential decrease in secondary particle, and thus spallation isotope, density away from the muon track.  $F_q$ increases due to excess energy loss producing more secondary particles and spallation isotopes.  Their functional forms are given in Refs.~\cite{Hosaka2006,Ishino}.

The likelihood function in Eq.~(\ref{eq:solar}) does not directly reflect shower physics.  The $Q_\mathrm{res}$ variable includes energy loss from all showers and low-energy delta rays along a muon track.  It can be close to the energy of the largest shower if that shower is very energetic, but most commonly it sums over comparable small showers and low-energy delta rays.  In addition, this likelihood does not include shower position information.

A new likelihood function Super-K recently developed, which is applied to the DSNB analysis~\cite{Bays2012}, is
\begin{equation}
\label{eq:dsnb}
\mathcal{L}_\mathrm{peak} = F_t(t)\cdot F_l(L_\mathrm{trans})\cdot F_q'(Q_\mathrm{peak})\cdot F_l'(L_\mathrm{long}) .
\end{equation}
Here $Q_\mathrm{peak}$ is the total light in the central 4.5 m of the reconstructed peak and $L_\mathrm{long}$ is the longitudinal distance (along the muon track) between the peak and the candidate event.  $F_l'$ deceases with respect to the absolute value of $L_\mathrm{long}$ because spallation isotopes are most frequently produced in the biggest showers.  $F_q'$ increases because secondary particle path lengths, and hence spallation production, are proportional to shower energy.  Their functional forms are given in Ref.~\cite{Bays}.

The likelihood function in Eq.~(\ref{eq:dsnb}) improved upon that in Eq.~(\ref{eq:solar}) and reflects some shower physics, though this was not recognized as the reason.  The variable $Q_\mathrm{peak}$ would be a good measure of the shower energy if the reconstruction were perfect, but it is not accurate, as we explained earlier.  This likelihood keeps the peak position of the biggest shower from reconstruction but discards the shape of the shower and smaller showers.

It is easy to see how our work on improving muon light profile reconstruction improves the efficiency of this likelihood.  First, we can measure the true shower energy without overcounting or undercounting problems (see Sec.~\ref{sec:improvC}).  Second, the functional form of $F_l'$ gets sharper (see Fig.~12 of Ref.~\cite{Li2015}), which more clearly separates spallation decays from random events.

To fully utilize the information about showers in the reconstructed muon light profiles, we propose a new likelihood,
\begin{equation}
\label{eq:profile}
\mathcal{L}_\mathrm{shower} = F_t(t)\cdot F_l(L_\mathrm{trans})\cdot F_z(z).
\end{equation}
Here $z$ is the position along the muon track from where it enters the ID.  For a muon with a faithfully reconstructed shower profile, $F_z(z)$ is the shower profile intensity at position $z$.  We emphasize that this shower profile is the one shown in Fig.~\ref{reconstruction} with Improvements $1+2+3$, which excludes the light from a minimum-ionizing muon and some low-energy delta rays.

The likelihood function in Eq.~(\ref{eq:profile}) fully incorporates information about showers.  Spallation production at each position is roughly proportional to the local secondary particle path length, which is roughly proportional to the local light intensity.  Random events have a flat position distribution along the muon track.  $F_z$ directly reflects how the probability of spallation production varies along a particular muon track.

\begin{figure}[t]
\begin{center}                  
\includegraphics[width=\columnwidth]{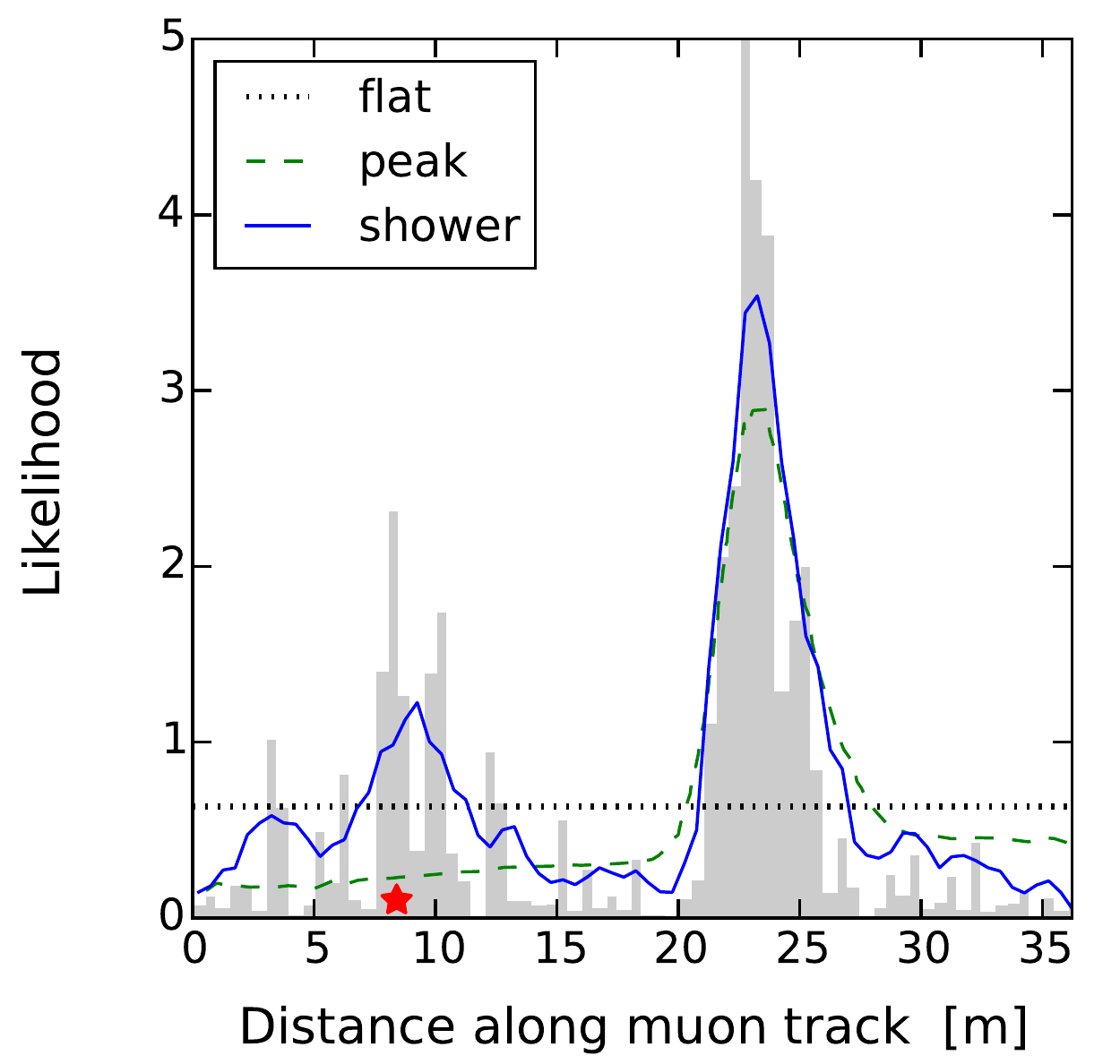}
\caption{Comparison of the $L_\mathrm{long}$ or $z$ component of the three different likelihood functions, normalized to equal area.  The $y$ axis unit is the Cherenkov light intensity, but we also use it for likelihoods because they have arbitrary normalization.  The gray shaded region is the real muon light profile, with muon light subtracted.  (We cut off the $y$ axis at 5, but the light intensity at 23 m goes to 8.)  The red star at 8 m indicates the position of a spallation event.}
\label{likelihood}
\end{center}
\end{figure}

Figure~\ref{likelihood} illustrates the term containing $L_\mathrm{long}$ or $z$ in these likelihood functions for an example muon.  The real muon light profile is shown in gray.  There are two showers.  Most frequently, the isotope would be associated with the larger one.  However, in this case, the isotope is produced in the smaller shower.

To emphasize the shape differences of these three likelihood functions, we normalize them to the same area.  $\mathcal{L}_\mathrm{flat}$ cuts background equally everywhere along the muon track.  It cuts events in shower regions too weakly and nonshower regions too strongly.  $\mathcal{L}_\mathrm{peak}$ correctly picks out the biggest shower along the muon track and cuts events in that region with more weight, although its shape does not trace this shower perfectly because the likelihood is from an average shower profile.  Further, when there are multiple showers, $\mathcal{L}_\mathrm{peak}$ cuts events in the small-shower regions too weakly. For $\mathcal{L}_\mathrm{shower}$, the likelihood function is the reconstructed shower profile itself.  As shown in this example, it not only traces the large shower well, but it also picks out the small shower.  It cuts spallation isotopes with weight proportional to the local shower intensity, which is close to optimal.

\subsection{Efficiency Improvements}
\label{sec:B}

We now quantify the improvement of our shower likelihood function over the Super-K likelihood functions.  To do so, we need to explain how to apply a cut, i.e., decide whether to discard a candidate event as a spallation decay on the basis of a likelihood test.

When applying a cut to a candidate event, one does not know which muon, if any, is correlated with it.  This is different from building a likelihood, where every spallation decay is paired with its parent muon and every random event is paired with an uncorrelated muon.

The first step of applying a cut is thus to build an event likelihood $\mathcal{L}_C(C)$ that returns the likelihood of an event being a spallation decay (from any muon).  This is done by taking a likelihood function $\mathcal{L}(C,M)$ and marginalizing over all muons $\{M_i\}$ that are possibly correlated with this candidate, which in practice are muons in the previous 100 s ($\sim 200$ muons).  One calculates $\mathcal{L}_i=\mathcal{L}(C,M_i)$ for each muon.  The maximum value of $\mathcal{L}_i$ is then assigned to this candidate event as its event likelihood:
\begin{equation}
\mathcal{L}_C(C) = \max_i \mathcal{L}(C,M_i) .
\end{equation}

Second, one obtains a spallation decay event sample and a random event sample, and finds the distributions of $\mathcal{L}_C$ for both samples.  The methods to get event samples are as described in Sec.~\ref{sec:A}, except that one discards the information about the candidate-muon correlations.  Then, one calculates the distributions of the event likelihoods $\mathcal{L}_C$.

For a specific likelihood function, the distributions of $\mathcal{L}_C$ for the spallation and random samples both have a bump, and drop off at small and large values.  By design, the average $\mathcal{L}_C$ value for the spallation sample is larger than that for the random sample.  However, the two distributions have significant overlap, which is why it is difficult to categorize a candidate as a spallation decay or a random event.  Because of the arbitrary normalization of $\mathcal{L}(C,M)$, one should not compare $\mathcal{L}_C$ distributions for different likelihood functions.

Third, one chooses a value, $\mathcal{L}_\mathrm{cut}$, that best separates the $\mathcal{L}_C$ distributions from the two samples.  The choice of $\mathcal{L}_\mathrm{cut}$ determines its effects on the signal and backgrounds, characterized by deadtime and cut efficiency.  Deadtime is the fraction of random events with $\mathcal{L}_C>\mathcal{L}_\mathrm{cut}$, which defines the signal loss.  Cut efficiency is the fraction of spallation decays with $\mathcal{L}_C>\mathcal{L}_\mathrm{cut}$, which describes the background rejection.  Hence, we want to minimize deadtime while we maximize cut efficiency.  For too small a value of $\mathcal{L}_\mathrm{cut}$, the deadtime would be unacceptably high.  For too large a value of $\mathcal{L}_\mathrm{cut}$, the cut efficiency would be unacceptably low.  An optimal value $\mathcal{L}_\mathrm{cut}$ must be chosen to maximize the signal detection significance.  For Super-K flat likelihood function, the cut efficiency and deadtime for the optimal $\mathcal{L}_\mathrm{cut}$ are about 90\% (10\% background remaining) and 20\% for their solar neutrino analyses~\cite{Ishino,Hosaka2006}. 

Finally, one applies the cut to the real data sample, rejecting events with $\mathcal{L}_C > \mathcal{L}_\mathrm{cut}$.

Now we can compare the efficiencies of different likelihood functions.  We vary the $\mathcal{L}_\mathrm{cut}$ value for each likelihood, obtaining pairs of deadtime and cut efficiency values.  The optimal $\mathcal{L}_\mathrm{cut}$ value would correspond to specific values.

\begin{figure}[t]
\begin{center}                  
\includegraphics[width=\columnwidth]{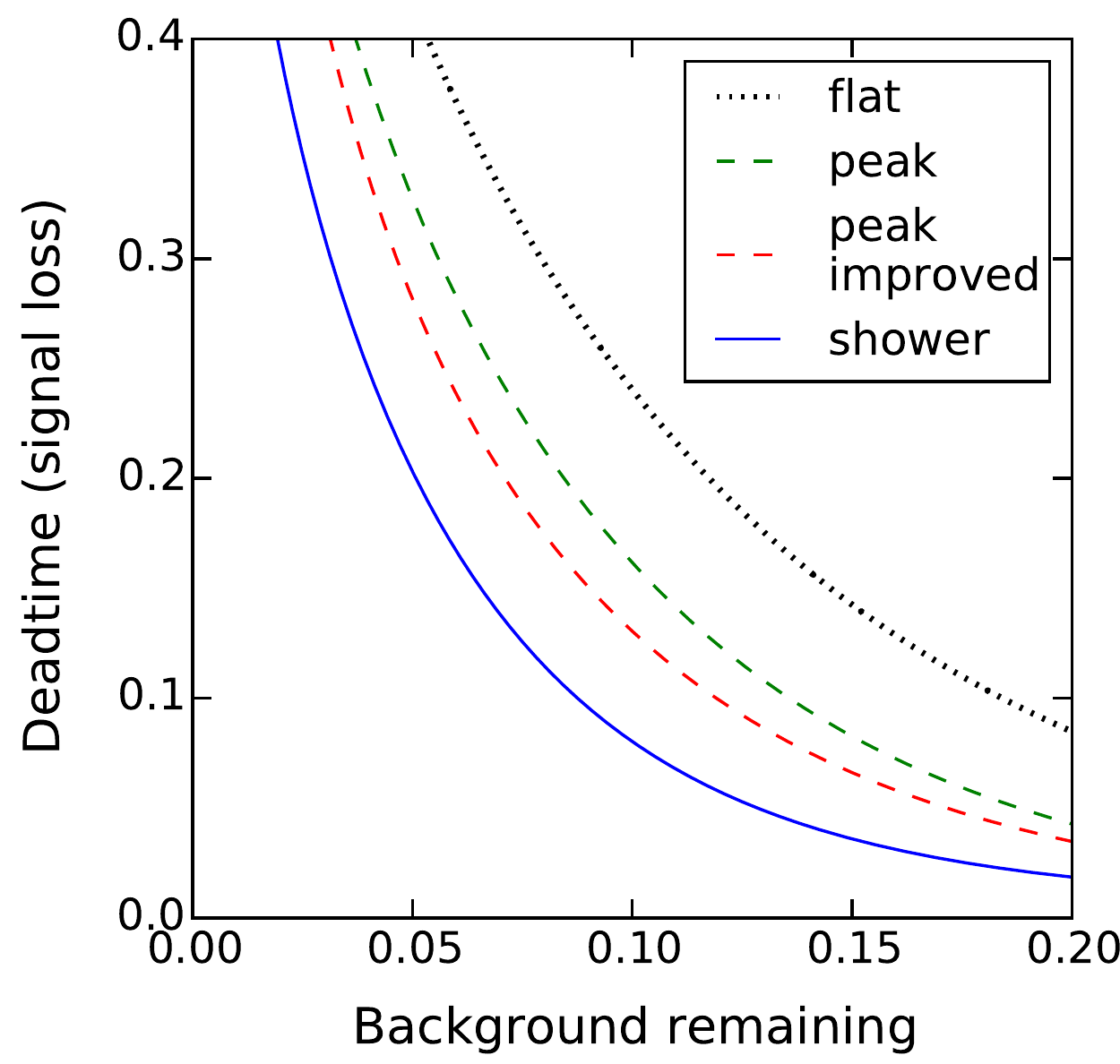}
\caption{Efficiency comparison for different likelihood functions.  The $x$ axis shows the fraction of spallation background events remaining after cuts.  The $y$ axis shows the fraction of signal events rejected by the cuts.}
\label{efficiency_dt}
\end{center}
\end{figure}

To better separate the factors that contribute to the differences between likelihood functions, we make some simplifications.  First, to show the maximum improvement possible due to better reconstruction methods, we take the real (simulated) muon light profiles instead of the reconstructed ones.  Second, to fairly compare the difference between the peak and shower likelihood functions, we add an improved peak likelihood that we explain below.  Third, we include only single throughgoing muons.  Lastly, our spallation samples have only spallation decays, whereas in Super-K analyses there are some random events.  Despite these simplifications, our results for the flat likelihood are consistent with the Super-K measurements and the differences among different likelihoods should be reasonably accurate.

Figure~\ref{efficiency_dt} shows the deadtime and cut efficiency for the three likelihood functions.  To emphasize the improvement, we show background remaining, which is $(1- \mathrm{cut\, efficiency})$, on the $x$ axis.  The flat [Eq.~(\ref{eq:solar})], peak [Eq.~(\ref{eq:dsnb})], and shower [Eq.~(\ref{eq:profile})] likelihoods take the functional forms we defined, with components taken from Super-K measurements or our definitions.  To take into account the fact that we use real muon light profiles, we show the peak improved likelihood, which is based on the Super-K peak likelihood formula, but we adjust $F_l'(L_\mathrm{long})$ (both shown in Fig.~12 in Ref.~\cite{Li2015}), assuming that the peak position can be measured perfectly.  

To make the comparisons specific, we compare likelihoods at fixed deadtime ($\simeq 20\%$), which focuses on background reduction.  (One could also compare at fixed cut efficiency, which focuses on signal gain.)

Going from the flat likelihood to the Super-K peak likelihood, the background remaining decreases from 0.12 to 0.09.  This shows how the method of Ref.~\cite{Bays2012} used for the DSNB analysis could benefit Super-K solar studies by focusing cuts on regions where the muon light profile peaks.  With a better peak localization, as our techniques could provide, the improvement would be to 0.07.  Finally, with our new shower likelihood, the complete improvement would be from 0.12 to 0.05.  Thus, it should be possible to reduce backgrounds by more than a factor of 2 with the results we present in this paper.  This can be combined with other cuts we have suggested or will present in forthcoming papers.

Our proposed new shower likelihood function, based on better reconstructed muon light profiles, could substantially reduce backgrounds.  It takes variables directly from each muon light profile, so it should be easy to implement.

In addition to the single throughgoing muons we consider, there are three other classes of muons identified by Super-K that are subdominant but relevant~\cite{Bays2012}.  Stopping muons only make isotopes when $\mu^-$ undergo nuclear capture~\cite{Galbiati2005,Li2015}, and we discuss cuts in Ref.~\cite{Li2015}.  Corner-clipping muons are just a category of throughgoing muons where reconstruction is more difficult.  Isotopes will be produced in the FV only when there is a shower that enters the ID (excess light in the outer detector may help identify large showers), because the lateral extent of a shower ($\lesssim 1$ m) and thus of isotope production is less than the thickness of the ID-FV shielding (2 m)~\cite{Li2014,Li2015}.  For multiple muons, also known as muon bundles, pairs of muons produced in the same atmospheric shower are the most common case~\cite{Hong1994,Becherini2006,Grassi2014}.  Higher-multiplicity events can be cut aggressively without appreciable deadtime.  Our reconstruction method could be adapted to deal with pairs, treating them together when the separation is $\lesssim 1$ m and singly when it is larger, along with straightforward adjustments for the amount of light and number of showers expected.  In summary, we see no barriers to adapting our methods to implement a complete background-rejection program in Super-K.


\section{Conclusions}
\label{sec:conclusion}

Muon-induced spallation backgrounds are a major source of background for low-energy neutrino detection.  Nevertheless, a complete picture of how these spallation isotopes are produced and a strong enough background rejection method have been lacking.  We are conducting a series of studies intended to provide a comprehensive theoretical understanding of how muons make spallation isotopes and to propose better ways to reject them.

In our previous papers~\cite{Li2014,Li2015}, we found that almost all spallation isotopes are produced by secondary particles, and that almost all secondary particles are made in muon-induced showers.  Our calculations agree with Super-K measurements on the total spallation yield and other data.  We also explained an empirical cut that Super-K developed for their DSNB analysis~\cite{Bays2012}.  However, one discrepancy remained: The Super-K reconstructed muon light profiles show prominent bump features, which are grossly inconsistent with shower physics.

In this paper, we show that the observed bump features are indeed caused by muon-induced showers.  The reason that they look too wide and short compared to showers can be traced back to the Super-K muon light profile reconstruction method.

We suggest ways to improve the Super-K reconstruction method.  By measuring the position and time for every PMT hit, Super-K solves a quadratic equation for the emission position of the photon along a muon track.  However, due to the electron deflection in showers and detector resolution, the quadratic equation could have zero, one, or two solutions for one PMT hit, and the solutions could be shifted from their true values.  We propose ways to improve this by picking out one solution when there are two, recovering one solution when there seem to be zero, and focusing on the PMT hits that give solutions closest to the true values.  We show that our improvements could lead to almost perfectly reconstructed muon light profiles.

We then propose a new spallation likelihood function, based on better reconstructed methods, that fully exploits the information contained in muon light profiles.  We demonstrated that it, combined with a better reconstruction method, can reduce the remaining background by a factor of 2 compared to the Super-K DSNB analysis cut, and even more compared to their solar analysis cut.

Our results could be easily adopted by Super-K for their solar neutrino and DSNB analyses.  The background rejection improvement could be especially dramatic for solar neutrinos, where the current cut does not even take advantage of the muon light profile peaks, much less the full understanding of shower physics.

The techniques we developed will benefit other neutrino experiments.  Our results have immediate applications for other water Cherenkov detectors, e.g., Hyper-Kamiokande~\cite{Abe2011hk}, which will be shallower than Super-K.  And, because our reconstruction method does not depend on the geometry of the PMTs, it could be applied to muon reconstruction in high-energy neutrino telescopes like IceCube~\cite{Aartsen2015nss}, where fluctuations in shower energy along muon tracks are used to estimate muon energy.  Finally, our reconstruction technique does not depend on the direction of the light, so our results could be adapted for scintillator detectors [see Eq.~(\ref{eq:hyperbola}) for isotropic light emission], especially in large-scale next-generation detectors such as JUNO~\cite{An2015jdp}.

\section*{Acknowledgments}
S.W.L. and J.F.B. are supported by NSF Grant No. PHY-1404311 to J.F.B.  We are grateful to Kirk Bays, Stephen Brice, Mauricio Bustamante, Masayuki Nakahata, Kenny Ng, Eric Speckhard, Mark Vagins, and especially Michael Smy and Michael Wilking for helpful discussions.

%


\newpage

\appendix*

\section{Additional reconstruction examples}

Here we show examples of muon light profile reconstruction for a variety of other cases.

\begin{figure*}[hbp!]
    \begin{center}                  
        \includegraphics[width=246.0pt]{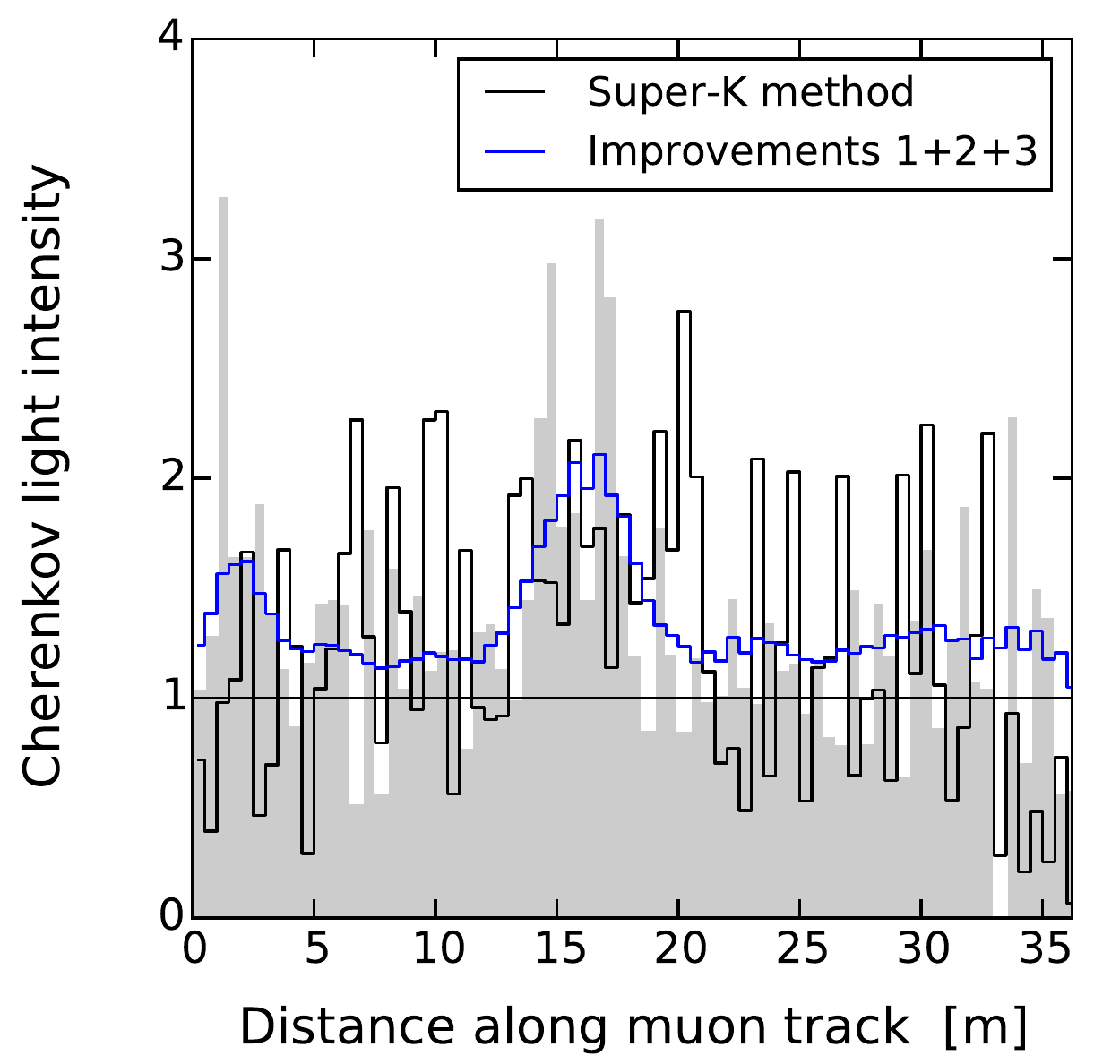}
        \hspace{0.25cm}
        \includegraphics[width=246.0pt]{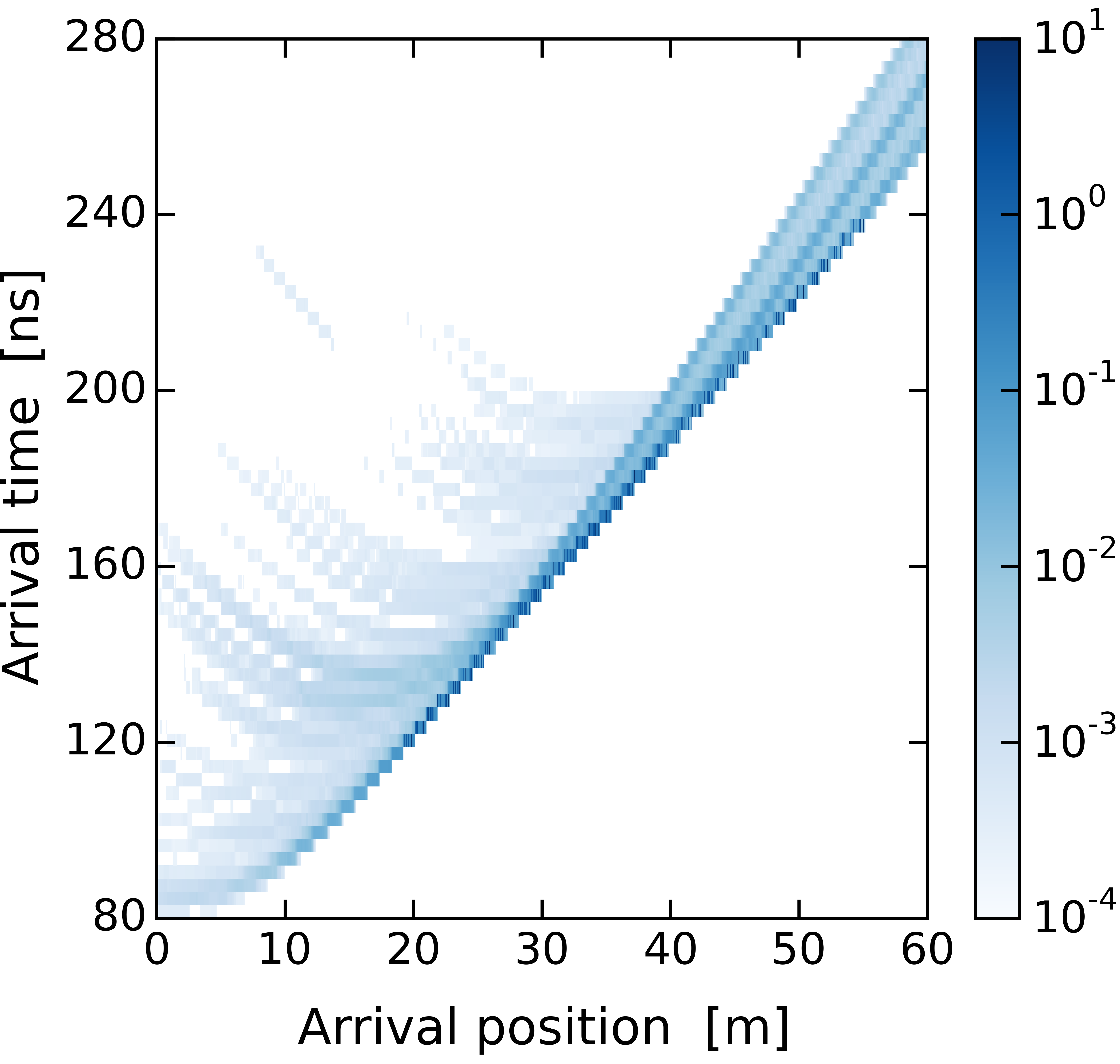}
        \caption{Shower reconstruction example---only small showers.  {\bf Left panel}: The gray shaded shape is the real (simulated) light profile.  The black line is the result of the reconstruction using the Super-K method.  The blue line is the result of our improvements.  {\bf Right panel}: The time and position pattern of Cherenkov light on the ID walls for this muon event.}
        \label{example_1}
    \end{center}
\end{figure*}

Figure~\ref{example_1} shows a muon event with only small showers.  There are two showers, at $\sim 2$ and 16 m.  The Super-K profile using the Super-K method does not reveal either shower, and is mostly noise.  Our reconstructed profile, however, successfully picks out both showers. 

\begin{figure*}[htp!]
    \begin{center}                  
        \includegraphics[width=\columnwidth]{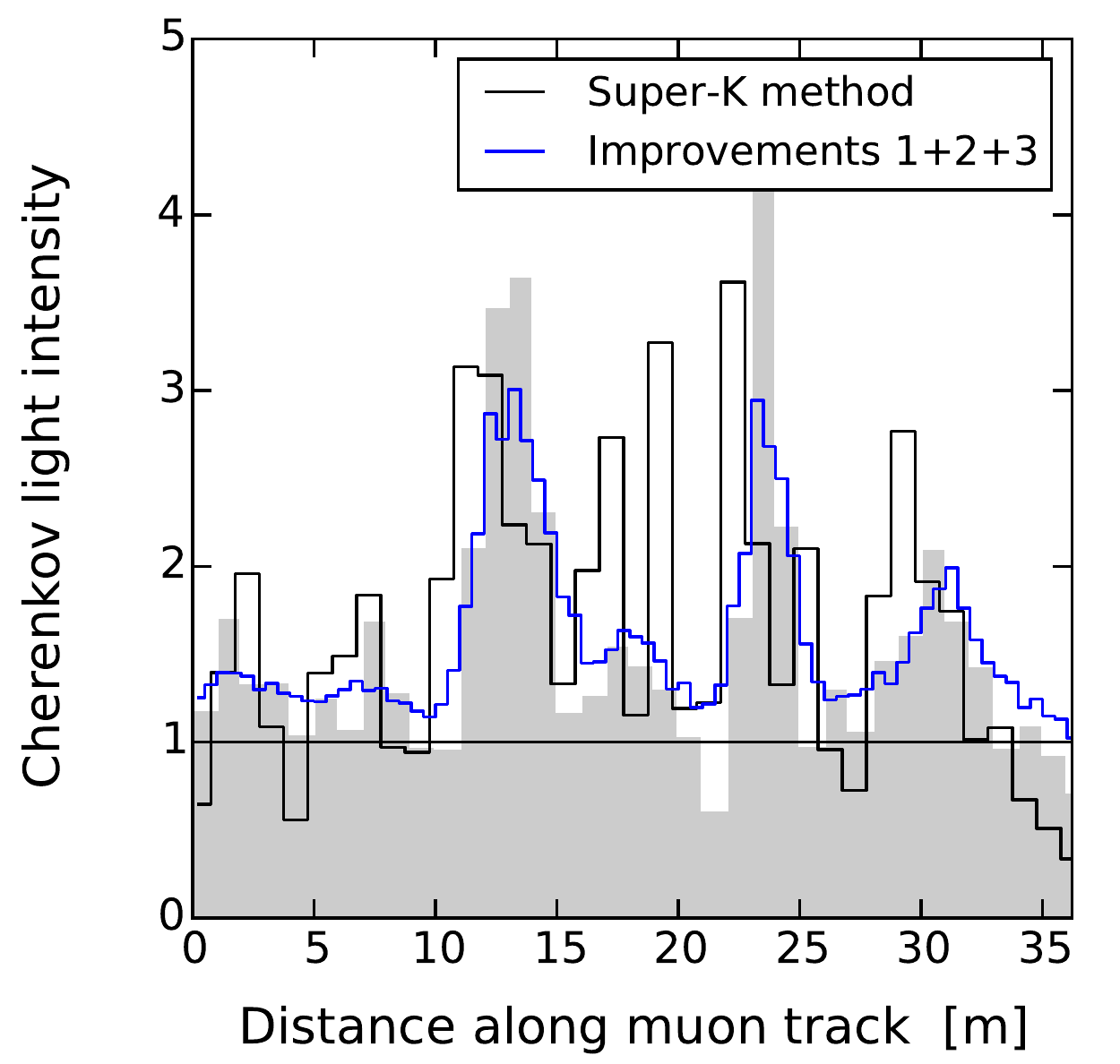}
        \hspace{0.25cm}
        \includegraphics[width=\columnwidth]{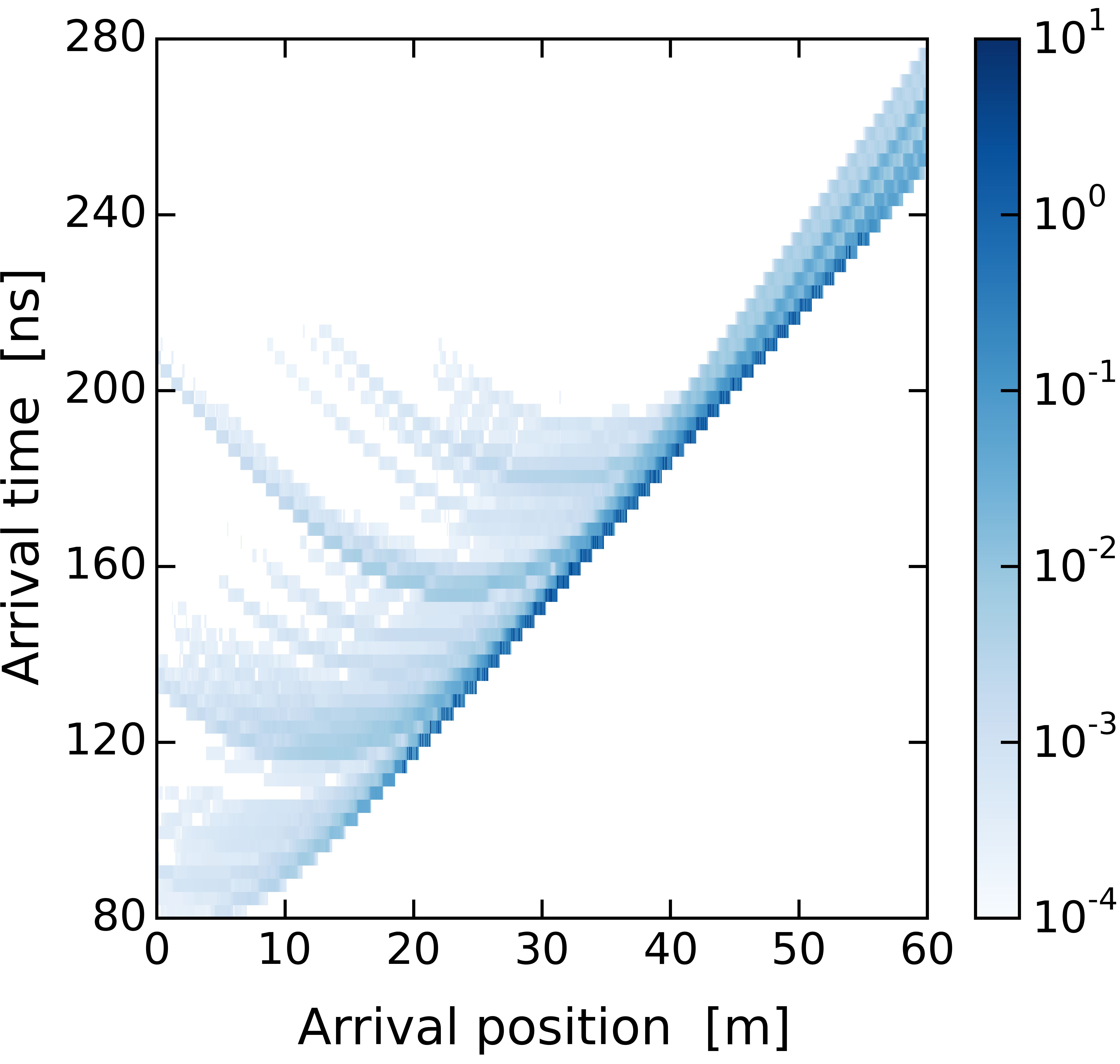}
        \caption{Shower reconstruction example---two comparable showers.  Descriptions as in Fig.~\ref{example_1}.  The only difference is that we use a bin size of 1 m in the left panel to reduce fluctuations.}
        \label{example_1052}
    \end{center}
\end{figure*}

Figure~\ref{example_1052} shows a muon event with moderate but comparable showers, at $\sim 12$ and 23 m (and smaller ones at $\sim 2$ m and 30 m).  The Super-K reconstructed profile is reasonable for one shower, poor for the others, and has a false shower in the middle.  Our reconstructed profile reconstructs all four showers clearly. 

\begin{figure*}[hbp!]
    \begin{center}                  
        \includegraphics[width=\columnwidth]{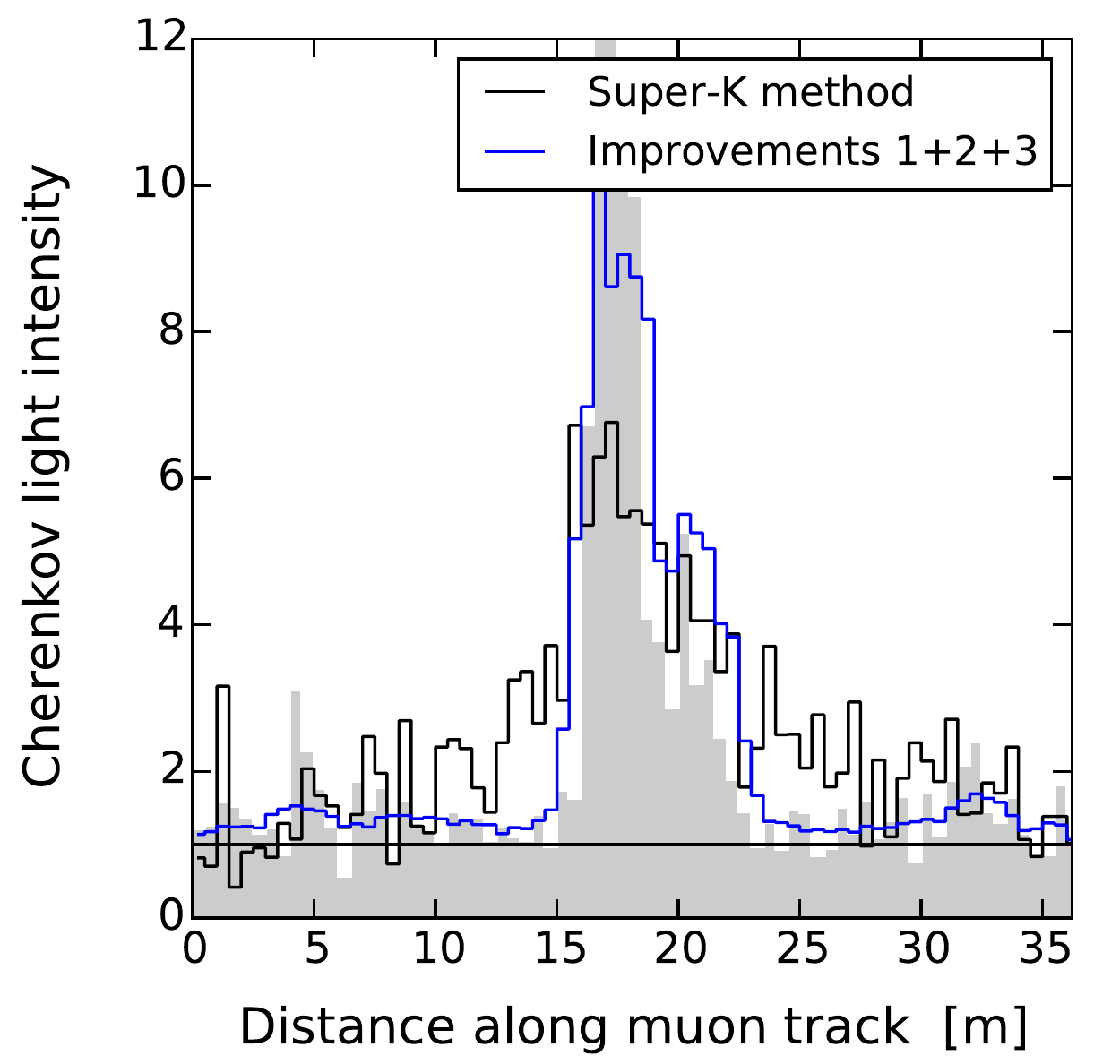}
        \hspace{0.25cm}
        \includegraphics[width=\columnwidth]{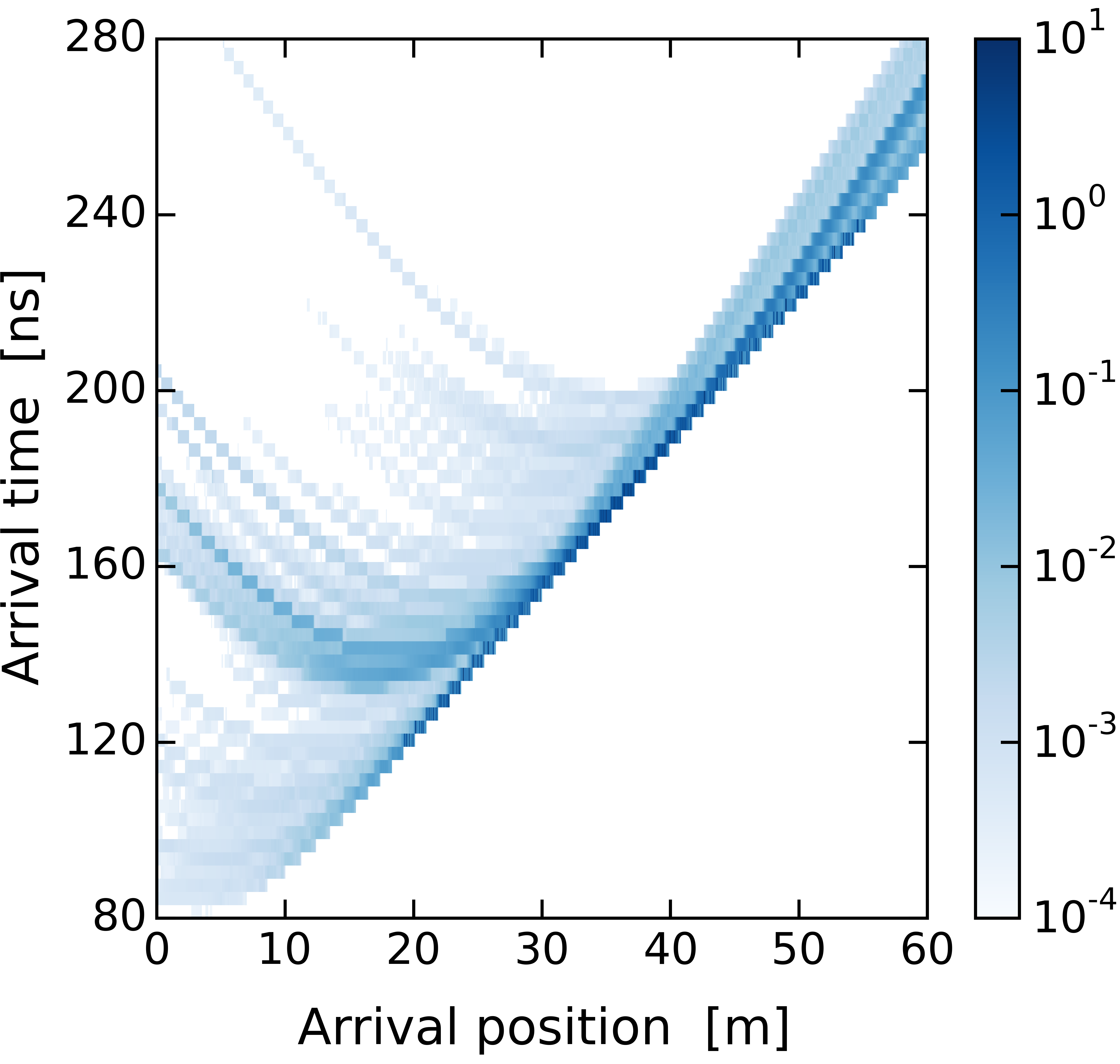}
        \caption{Shower reconstruction example---hadronic showers.  Descriptions as in Fig.~\ref{example_1}}
        \label{example_108}
    \end{center}
\end{figure*}

Figure~\ref{example_108} shows a muon event with a hadronic shower.  This is a relatively clean event.  Both profiles have the correct peak position, and trace the shower shape.  The Super-K profile is more smeared out.  Our reconstructed shower is not as sharp as large in other examples due to the larger lateral displacement of charged particles in hadronic showers. 

In summary, these examples demonstrate the better performance of our reconstruction for small, multiple and hadronic showers.

\vspace{1.5em}

\onecolumngrid

\twocolumngrid

\end{document}